\documentclass[a4paper,10pt]{article}

\usepackage{latexsym}
\usepackage{amsmath,amsfonts,amssymb,amsthm}
\usepackage{graphicx,color,epsfig}
\usepackage{algorithmic}

\def\T{\mathbb{T}}
\def\R{\mathbb{R}}
\def\N{\mathbb{N}}
\def\Z{\mathbb{Z}}
\def\Q{\mathbb{Q}}

\newtheorem{Def}{Definition}
\newtheorem{Claim}[Def]{Claim}
\newtheorem{Lem}[Def]{Lemma}
\newtheorem{Cor}[Def]{Corollary}
\numberwithin{equation}{section}

\begin{document}

\noindent
{\LARGE \bf Computer-assisted proof of ergodicity breaking in expanding coupled maps}
\bigskip

\noindent
{\large \bf Bastien Fernandez}
\bigskip

\noindent
Laboratoire de Probabilit\'es, Statistique et Mod\'elisation, CNRS - Univ.\ Paris Denis Diderot - Sorbonne Univ., 75205 Paris CEDEX 13 France
\bigskip

\begin{abstract}
{\small From a dynamical viewpoint, basic phase transitions of statistical mechanics can be regarded as a breaking of ergodicity. While many random models exhibiting such transitions at the thermodynamics limit exist, finite-dimensional examples with deterministic dynamics on a chaotic attractor are rare, if at all existent. Here, the dynamics of a family of $N$ coupled expanding circle maps is investigated in a parameter regime where absolutely continuous invariant measures are known to exist. At first, empirical evidence is given of symmetry breaking of the ergodic components upon increase of the coupling strength, suggesting that breaking of ergodicity should occur for every integer $N>2$. Then, a numerical algorithm is proposed which aims to rigorously construct asymmetric ergodic components of positive Lebesgue measure. Due to the explosive growth of the required computational resources, the algorithm successfully terminates for small values of $N$ only. However, this approach shows that phase transitions should be provable for systems of arbitrary number of particles with erratic dynamics, in a purely deterministic setting, without any reference to random processes.}
\end{abstract}


\section{Introduction}
In statistical mechanics, the term {\em phase transition} originally refers to the emergence of multiple Gibbs states for the Hamiltonian associated with a collective system \cite{R69}, most notably accompanied with symmetry breaking, as in the Ising model. While this notion has been formulated for equilibrium measures, from the dynamical viewpoint, it conveys some breakdown of ergodicity in a related dynamical process \cite{vEvH84}. Using out-of-equilibrium procedures, such ergodicity failures have been rigorously proved for suitably designed Markov processes and probabilistic cellular automata (PCA) \cite{H00,L05}. More precisely, Metropolis rules, Glauber dynamics and the like have been designed to account for relaxation to pre-constructed equilibrium states, including the case when several such states coexist \cite{M81}.

Notwithstanding the success of this dynamical approach to phase transitions, the fact that classical mechanics is ruled by deterministic laws of motion calls for evidences in purely deterministic systems, irrespective of any consideration of random processes. Barring the use of out-of-equilibrium procedures, can chaotic attractors in deterministic analogues of random models of interacting particles systems exhibit ergodicity breaking (associated with symmetry breaking)? Despite having generated considerable attention, this question still eludes a satisfactory response that would exclude numerical flaws or unverifiable theoretical assumptions, even in basic examples such as networks of coupled expanding or hyperbolic maps.

Indeed, some examples of infinite lattices of interacting chaotic maps have been designed so that their dual dynamics acting on measures consist of PCA exhibiting phase transitions \cite{GM00,M05}. Thus, out-of-equilibrium approaches can in principle be lifted to deterministic dynamical systems. However, this operation requires explicit knowledge of a coupling-intensity-independent Mar\-kov partition, in particular one that ensures a pre-selected Markov chain/PCA for the dual measure evolution. Yet, such a complete understanding of Markovian properties is rare for realistic deterministic systems. This is especially the case of chaotic collective systems with homogenizing interactions \cite{B05}, which typically fail to fit the standard assumptions of the theory of dynamical systems, such as being diffeomorphisms. In general, very little is known about the symbolic dynamics, and, if at all, only for weak interaction intensity \cite{JP98,CF13}. This shortcoming calls for verifications of non-ergodicity in a purely deterministic setting, that are independent of any knowledge of the associated symbolic dynamics.

In this setting, various studies have reported changes in the global dynamics of coupled chaotic maps as their interaction strength is increased. Infinite lattice examples have been provided, for which the invariant densities associated with the Perron-Frobenius operator undergo bifurcations. In particular, an analogous transition to the one in the Curie-Weiss model of statistical mechanics, which only affects the dynamics at the thermodynamics limit, has been proved for the model in \cite{BKZ09}. Moreover, convergence to a point mass for strong coupling has been established in \cite{BKST18} for a mean-field model acting on measures on the circle. 

This convergence is reminiscent of phenomenological changes in finite systems. However, such alterations have always been observed to be preceded by a reduction in the Lyapunov dimension. Further, these changes have repeatedly resulted in stationary or periodic behaviors of the spatial averages of symmetry-related observables, see e.g.\ \cite{BBCFP95,CM93,J95,MH93}. This phenomenology is comparable with synchronization-like scenarios in which trajectories asymptotically shrink to lower dimensional subspaces. Hardly compatible with a coupling-independent symbolic grammar, they cast doubt on the nature of phase transitions in dynamical systems: do such transitions only take place at the thermodynamics limit? If not, do they necessarily require a lowering of the dimension of the attractor, or can they take place while hyperbolicity properties remain unchanged, and in absent knowledge of a Markov partition?

To address these issues, we consider a family of piecewise affine mappings $F_{N,\epsilon}$ of the $N$-dimensional torus $\T^N$, which mimic interacting particle dynamics driven by chaotic individual stirring and homogenizing interactions. Interactions are of mean-field type (all-to-all coupling) with adjustable strength $\epsilon$ (Section 2).\footnote{Details of the considerations in this paragraph and the following ones will be given below.}  Moreover, the mappings $F_{N,\epsilon}$ have many symmetries, most notably, they commute with flipping the sign of all coordinates. For $\epsilon=0$, the units are decoupled and evolve independently. For $\epsilon\in [0,\tfrac12)$, all mappings are expanding and must support (Lebesgue) {\bf absolutely continuous invariant measures} (ACIM). Moreover, each ACIM must be ergodic - and hence invariant under the action of any symmetry - when $\epsilon$ is small enough. For such mappings, it is therefore natural to 
(mathematically) examine ergodicity persistence or its failure in this expanding regime, together with accompanying symmetry features. 

Preliminary evidences, based on simulations of trajectories, are presented in Section \ref{S-EMPIRIC}. Completing previous findings in \cite{F14}, they reveal that for each $N>2$, ergodicity is broken via symmetry breaking, for $\epsilon$ sufficiently large. 
For $N=3$ and 4, this phenomenology has been confirmed by analytic demonstrations \cite{F14,S18,SB16}. For convenience, the proofs actually deal with $D=(N-1)$-dimensional mappings (denoted by $G_{D,\epsilon}$ below) whose ergodicity and its failure coincide with $F_{N,\epsilon}$. While these proofs seem to have captured essential information about the dynamics, their approach, which relies on inspiration from direct observations of trajectories in phase space, is hardly applicable when $D$ is large. 

In order to address large $D$ (or $N$), we propose to use instead computer-assisted proof. To that goal, an algorithm is introduced (Section 4) that aims to rigorously construct asymmetric $G_{D,\epsilon}$-invariant sets containing ACIM. The algorithm relies on empirical input from simulations and is based on two important properties.\footnote{Likewise, see the extended section 4.2 and additional information in the appendices for details and numerical implementation.} One characteristics is that the dynamics of suitable-for-the-proof polytopes can be encoded into one on vectors in $\R^{D(D+1)}$. In other words, the dynamics of suitable sets can be captured by a limited number of real variables. The other one is that computations can be designed so that, when $\epsilon$ is itself a rational number, they involve rational numbers only. 
Results of the algorithmic construction are given for $D$ up to 5 (Section 4.1). However, they clearly indicate that ergodicity breaking should be provable for arbitrary $D$. A discussion about possible improvements and alternative proofs is provided in Section 5.

\section{Expanding systems of piecewise affine globally coupled maps}
The mappings $F_{N,\epsilon}$ are defined as follows \cite{F14}
\[
(F_{N,\epsilon}(u))_i=2u_i+\frac{2\epsilon}{N}\sum_{j=1}^{N}g(u_j-u_i)\ \text{mod}\ 1,\ 1\leq i\leq N,
\] 
where $g$ represents pairwise elastic interactions on the circle \cite{KY10} and is defined by $g(u)=u-h(u)$ for all $u\in\T^1$ with
\[
h(u)=\left\{\begin{array}{ccl}
\lfloor u+\frac12\rfloor&\text{if}&u\not\in\frac12+\Z\\
0&\text{if}&u\in\frac12+\Z .
\end{array}\right.
\]
Here, $\lfloor \cdot \rfloor$ is the floor function. Hence $g$ is piecewise affine, with slope 1 and discontinuities at all points of $\frac12+\Z$. 


Mean field coupling implies $F_{N,\epsilon}\circ \sigma =\sigma\circ F_{N,\epsilon}$ for every $\sigma\in \Pi_N$, where $\Pi_N$ is the group of permutations of $\{1,\dots,N\}$. 
The symmetry $g(-u)=-g(u)\ \text{mod}\ 1$ implies commutativity with the {\bf sign flip} $-\text{\rm Id}_{\T^{N}}$. 

The mappings $F_{N,\epsilon}$ are non-singular. Therefore, considerations about ergodicity of their ACIM can ignore the dynamics of sets of vanishing Lesbegue measure, and in particular, the orbits of discontinuity sets.  
Away from these discontinuities, $F_{N,\epsilon}$ is piecewise affine with constant derivative
\[
(DF_{N,\epsilon}v)_i=2(1-\epsilon)v_i+\frac{2\epsilon}{N}\sum_{j=1}^{N}v_j,\ 1\leq i\leq N,
\]
from which it follows that $F_{N,\epsilon}$ is expanding for $\epsilon<\frac12$. As a consequence, 
its Milnor attractor \cite{Buescu97,M85} in this regime must consist of a finite union of Lebesgue ergodic components \cite{T01}, {\sl viz.}\ the attractor of almost every trajectory must be a set of positive Lebesgue measure (thereby excluding any dimension reduction). 

Focusing on this expanding regime, as mentioned above, we aim to address ergodicity of the attractor, that is whether the Lebesgue ergodic components are unique or not. Up to semi-conjugacy, this question can be examined in a more convenient family of piecewise affine mappings of the $D=(N-1)$-dimensional torus. The reduced mappings $G_{D,\epsilon}$ (defined below) have equivalent features to the originals. Namely, their symmetry group is isomorphic to $\Z_2\times \Pi_{D+1}$ and in particular, we have $G_{D,\epsilon}\circ -\text{\rm Id}_{\T^{D}}=-\text{\rm Id}_{\T^{D}}\circ G_{D,\epsilon}$. Furthermore, their asymptotic dynamics for $\epsilon\in [0,\frac12)$ must also lie in finitely many ergodic components of positive Lebesgue measure. 

More precisely, the transformation $\pi_N$ of $\T^N$ defined by \cite{SB16}
\[
(\pi_Nu)_i=\left\{\begin{array}{ccl}
u_i-u_{i+1}\ \text{mod}\ 1&\text{if}&1\leq i\leq N-1\\
\sum_{j=1}^Nu_j\ \text{mod}\ 1&\text{if}&i=N
\end{array}\right.
\]
(semi-)conjugates $F_{N,\epsilon}$ to the direct product $G_{N-1,\epsilon}\times F_{1,\epsilon}$ ({\sl viz.} $\pi_N\circ F_{N,\epsilon}=(G_{N-1,\epsilon}\times F_{1,\epsilon})\circ \pi_N$). The mapping $F_{1,\epsilon}(u)=2u\ \text{mod}\ 1$ (which acts on the sum coordinate $(\pi_Nu)_N$) does not depend on $\epsilon$ and is ergodic with respect to the Lebesgue measure on $\T^1$. Therefore, any failure of ergodicity for $F_{N,\epsilon}$ has to be concomitant with the same phenomenon for $G_{N-1,\epsilon}$.

The mapping $G_{N-1,\epsilon}$ does not involve the sum coordinate. Moreover, its (constant) derivative conveniently turns out to be a multiple of the identity in $\T^{N-1}$, {\sl i.e.}\ $DG_{N-1,\epsilon}=2(1-\epsilon)\text{\rm Id}_{\T^{N-1}}$. Explicitly, we have $G_{D,\epsilon}=2(1-\epsilon)\text{\rm Id}_{\T^{D}}+\tfrac{2\epsilon}{D+1}B_D\ \text{mod}\ 1$, where for $i\in\{1,\dots,D\}$
\[
(B_{D}(x))_i=2h(x_i)+\sum_{j=1}^{i-1}h(\sum_{k=j}^{i}x_k)-h(\sum_{k=j}^{i-1}x_k)+\sum_{j=i+1}^{D}h(\sum_{k=i}^{j}x_k)-h(\sum_{k=i+1}^{j}x_k).
\]

Perturbative arguments at the uncoupled limit $\epsilon=0$, applied to the related transfer operator acting on measure densities \cite{KK92}, demonstrate ergodicity for $\epsilon>0$ sufficiently small, for each integer $D$. Instead, the dynamics is not expanding at the limit $\epsilon=\tfrac12$. All mappings $G_{D,\frac12}$ consist of piecewise isometries and their global dynamics can be hardly described. In any case, continuation arguments appear inapplicable at this limit. In order to evaluate ergodicity or its failure at the upper end of the expanding regime, we therefore opted to collect numerical evidences. 

\begin{figure}[ht]
\begin{center}
\includegraphics*[width=27mm]{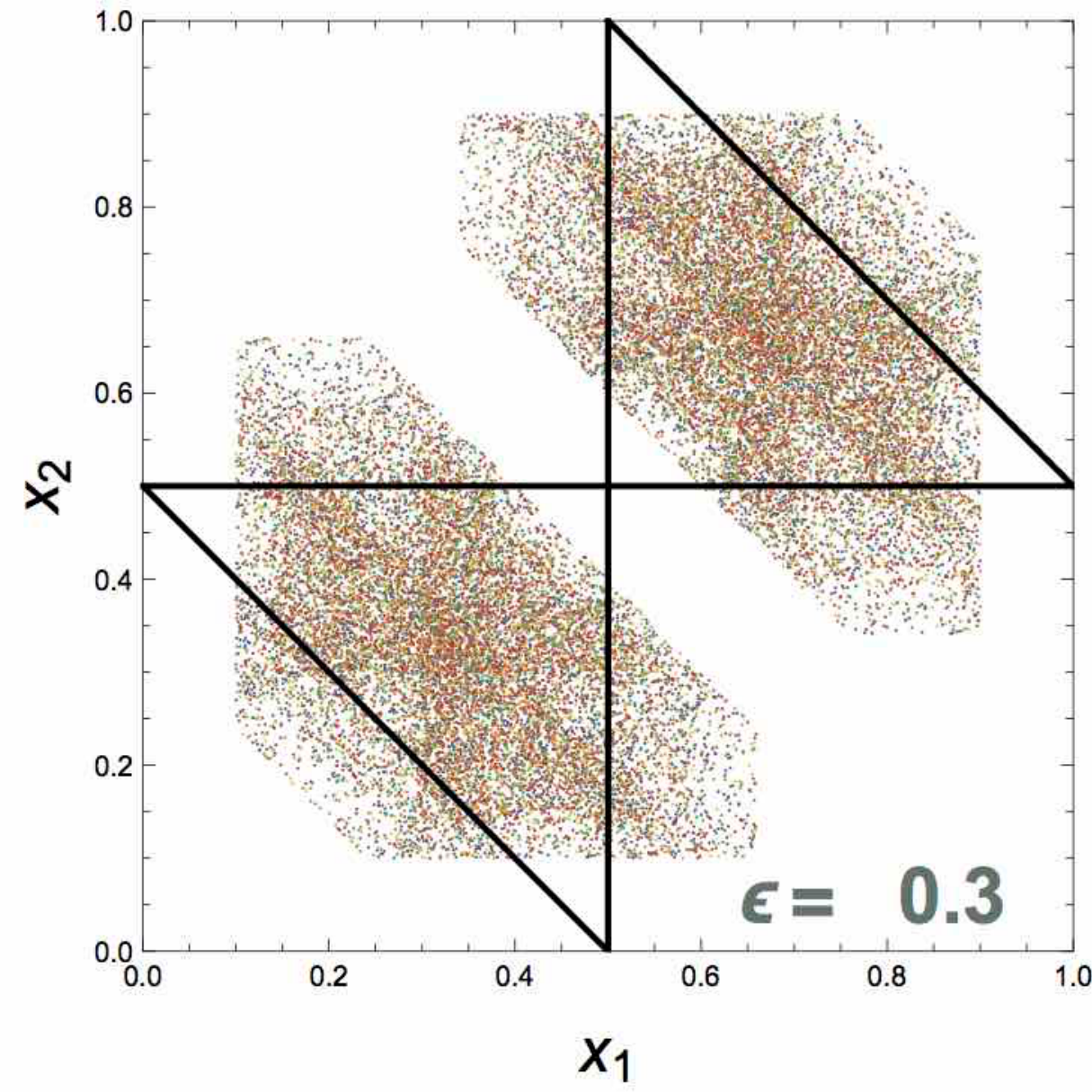}
\includegraphics*[width=27mm]{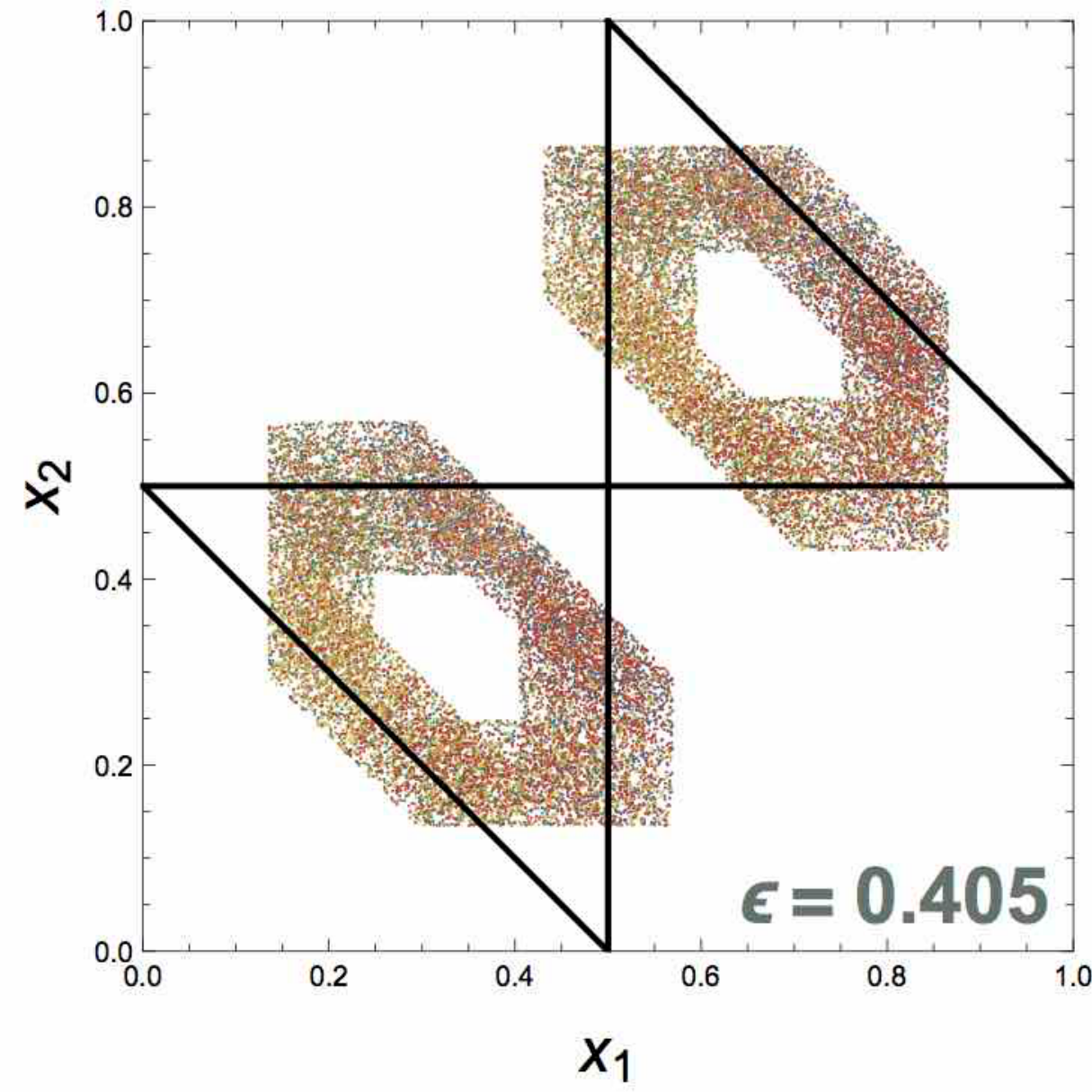}
\includegraphics*[width=27mm]{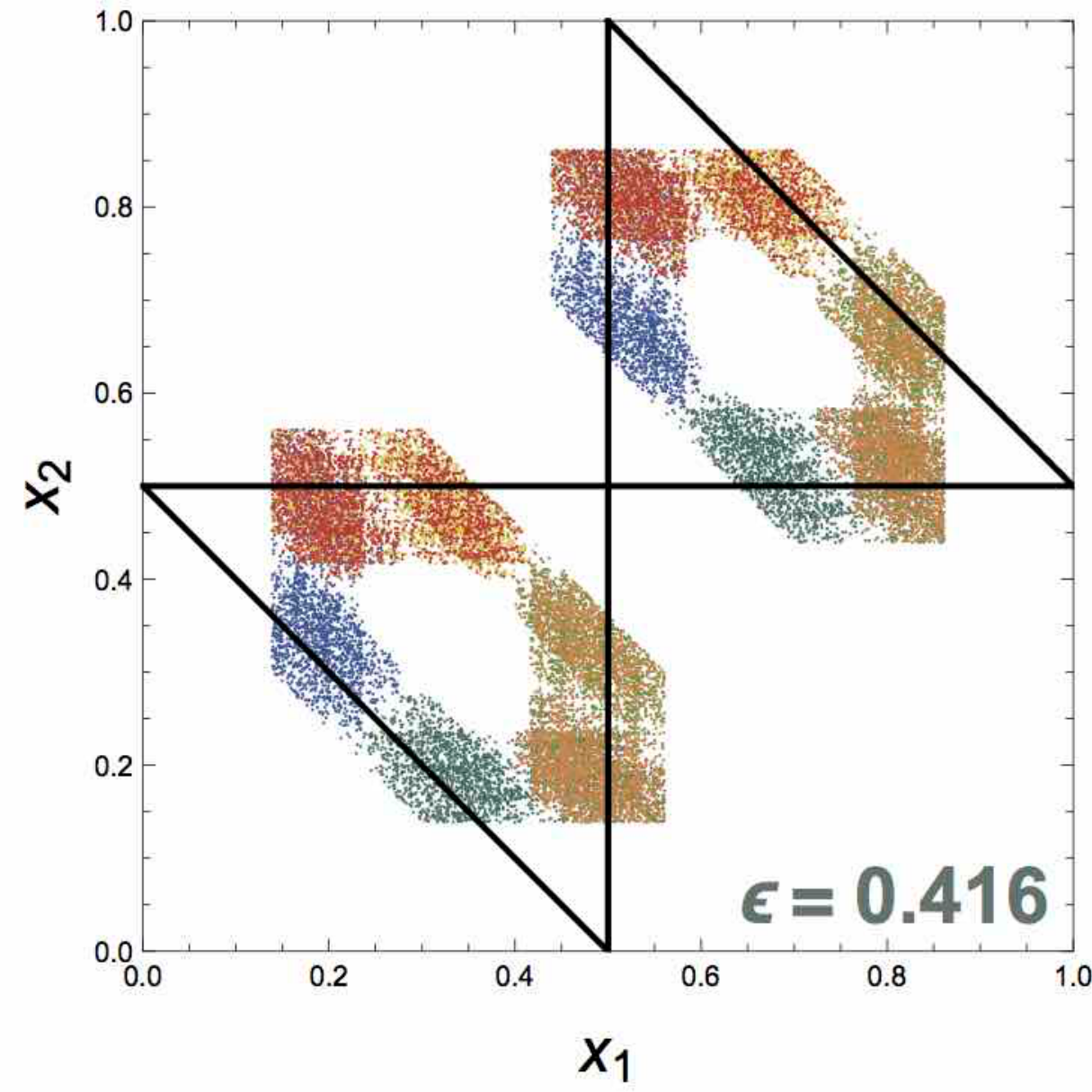}
\includegraphics*[width=27mm]{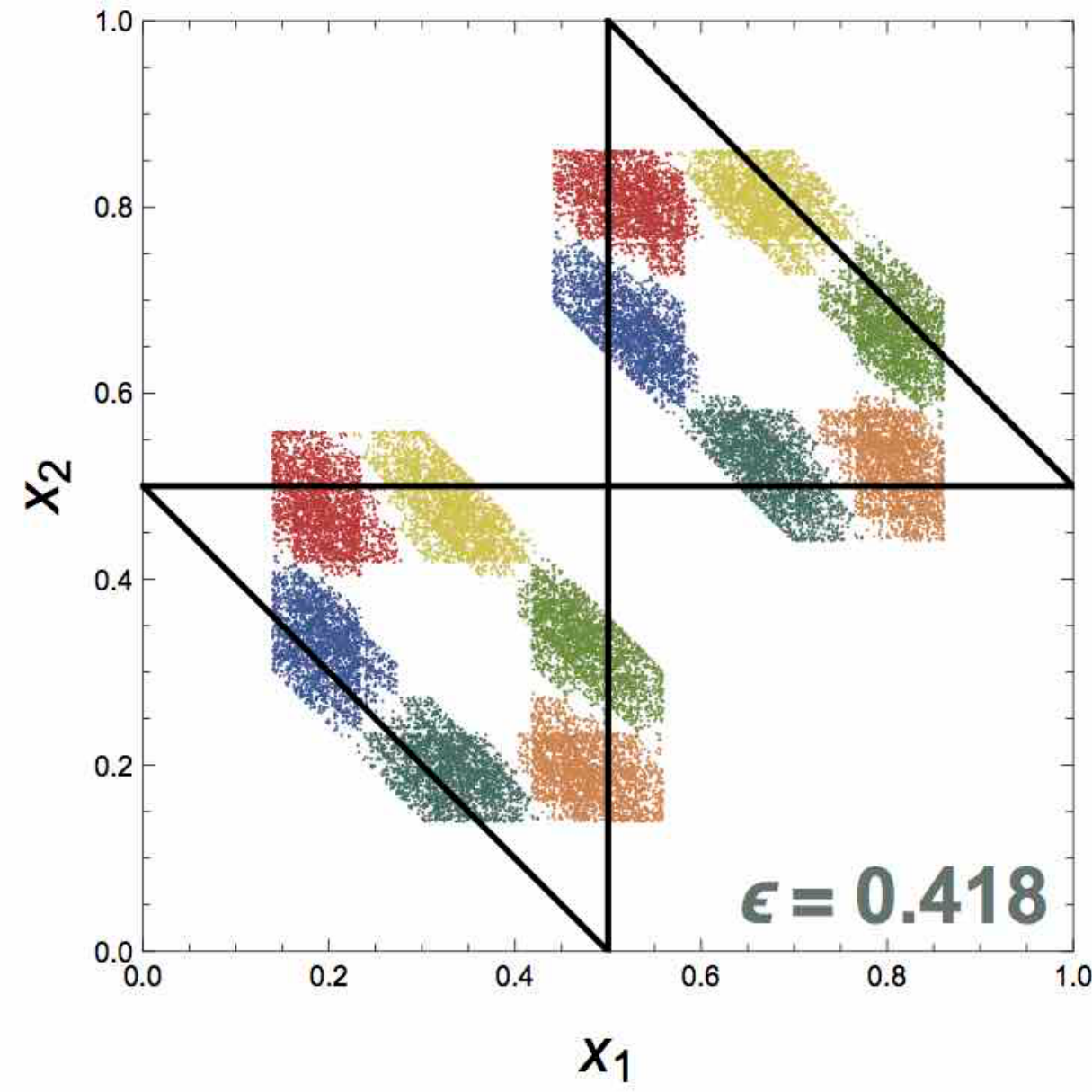}

\vspace{0.2cm}
\includegraphics*[width=27mm]{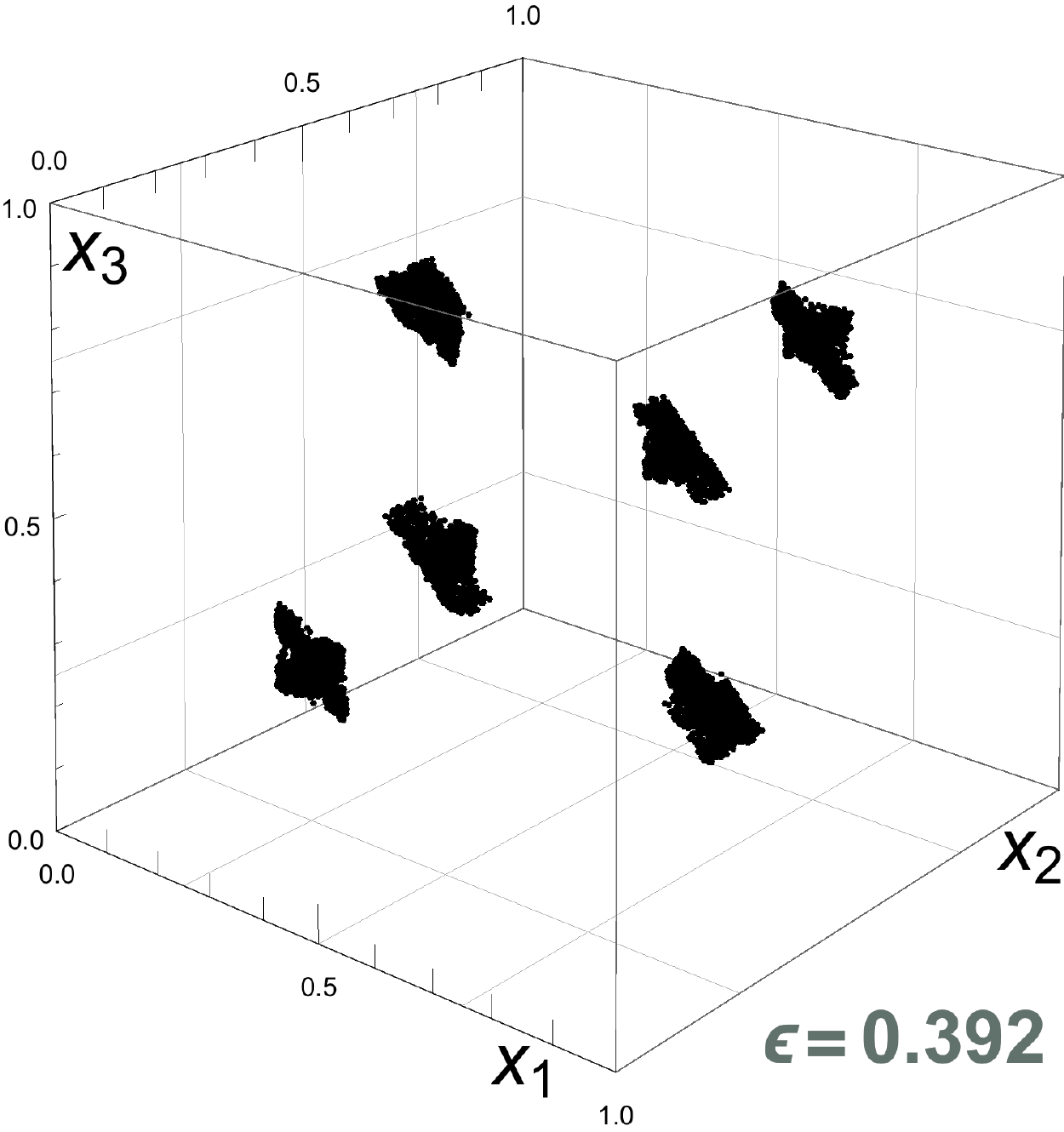}
\includegraphics*[width=27mm]{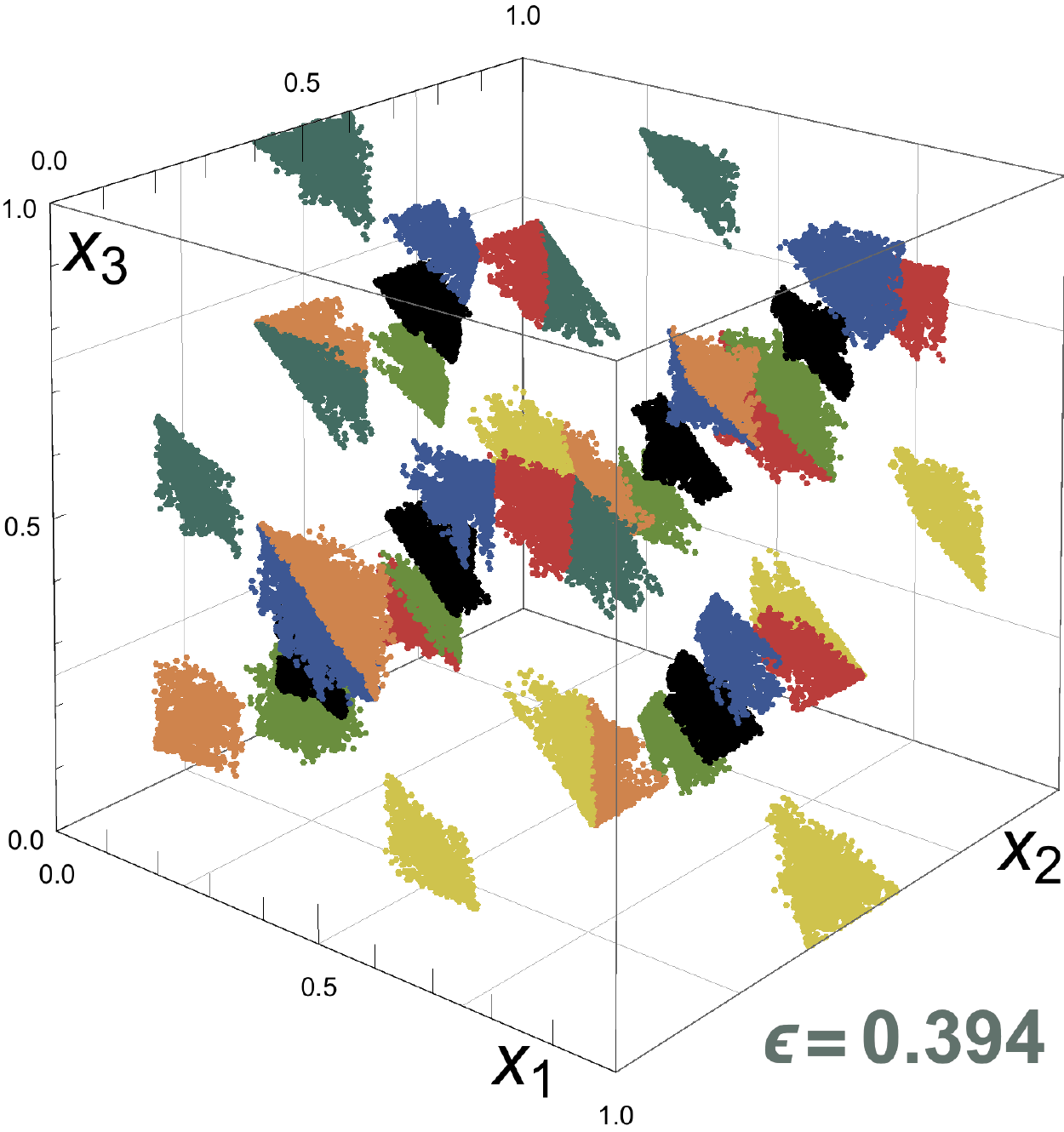}
\includegraphics*[width=27mm]{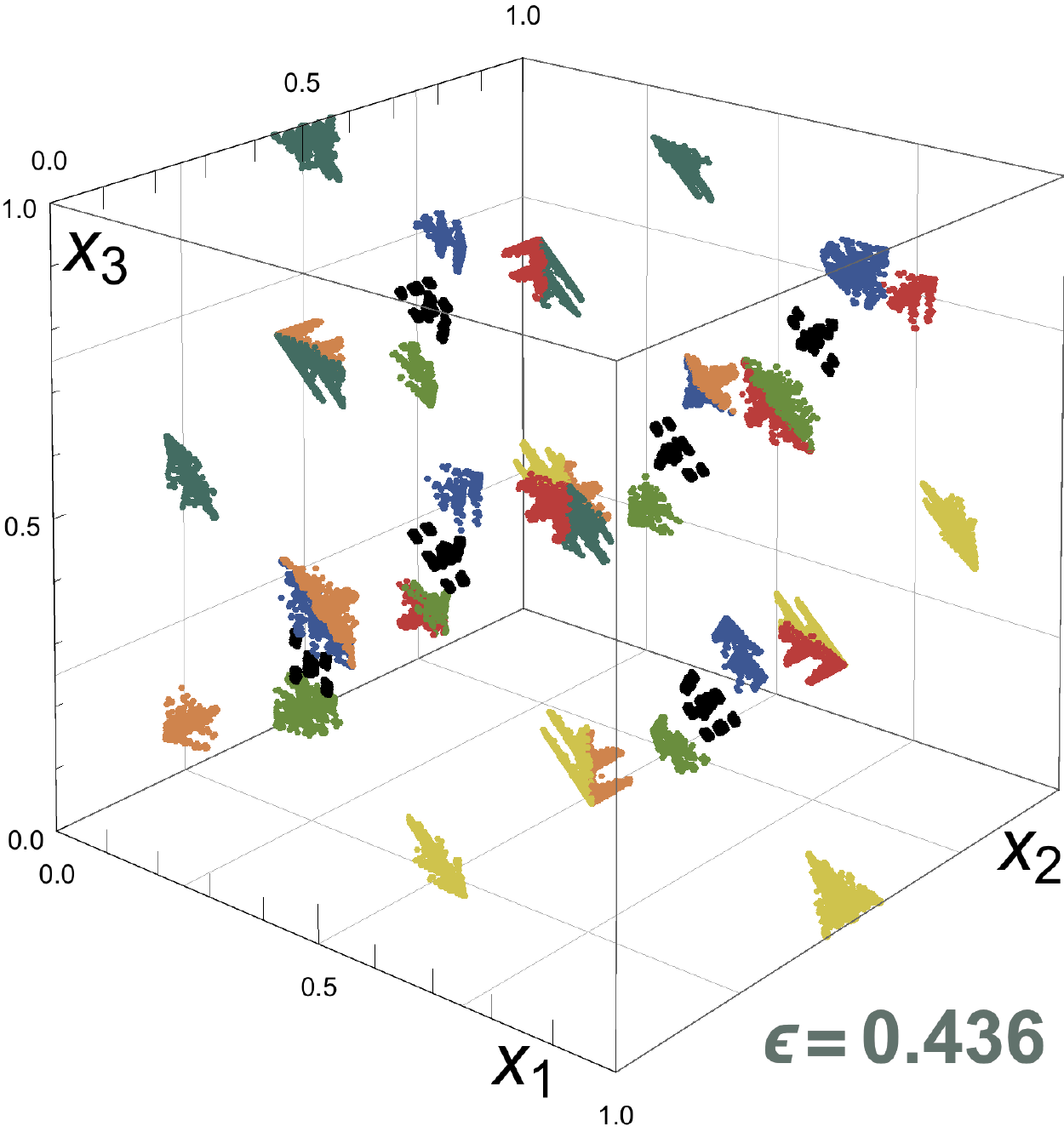}
\includegraphics*[width=27mm]{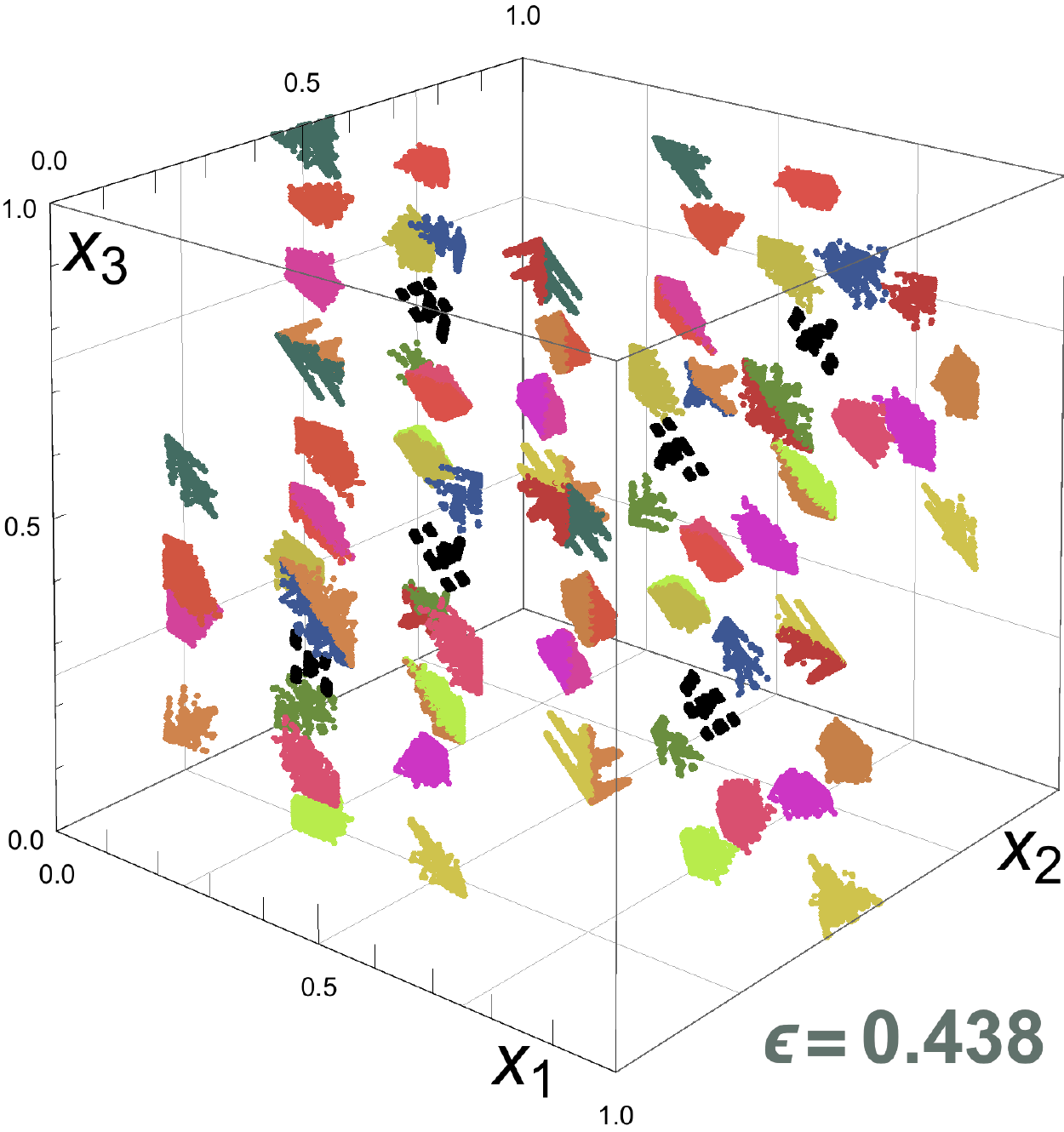}
\end{center}
\caption{Direct empirical evidence of ergodicity/symmetry breaking for the maps $G_{2,\epsilon}$ (top row) and $G_{3,\epsilon}$ (bottom row). Superimposed plots of $n_1$ consecutive points of orbits (one color for each orbit), after projection onto $(0,1)^D$ and discarding of the first $n_0$ iterates ($n_0=15\times 10^3$ and $n_1=5\times 10^3$ for $D=2$, $n_0=40\times 10^3$ and $n_1=10\times 10^3$ for $D=3$). In each image, one orbit is started from a representative initial condition and the other ones follow by applying symmetry. For $\epsilon<\epsilon_D$ (where $\epsilon_2\simeq 0.417$ and $\epsilon_3\simeq 0.393$), all orbits appear to cover the same set, suggesting that ergodicity holds. However, for $\epsilon>\epsilon_D$, no two points of distinct orbits overlap, suggesting the existence of multiple Lebesgue ergodic components. For $D=2$, discontinuity lines $x_{1,2}=\frac12$ and $x_1+x_2=\frac12,\frac32$ are also shown. For $D=3$, several color schemes are used to differentiate groups of distinct symmetry type: black for the symmetric trajectory and rainbow (resp.\ neon) colors for the 6 (resp. 8) element group, which exist for $\epsilon>\epsilon_3$ (resp.\ $\epsilon>0.437$).}
\label{PHASPA}
\end{figure}

\section{Empirical results from numerical trajectories}\label{S-EMPIRIC}
This section presents evidences of ergodicity breaking that were obtained from numerical simulations of trajectories. The hints are of two types: rendering of trajectories when $D\leq 3$ and order parameter estimates for arbitrary $D\geq 2$.\footnote{For $D=1$, no ergodicity failure occurs since the Milnor attractor of $G_{\epsilon,1}$ is transitive, and hence ergodic, for every $\epsilon\in [0,\frac12)$ \cite{F14,SB16}.} We begin by presenting evidences of the first type. 

In low dimension, direct visualization of asymptotic-orbit traces in phase space offers a straightforward evaluation of ergodicity/symmetry and their failure. On Fig.\ \ref{PHASPA}, late iterates of orbits started from representative initial conditions, and their images under symmetries, are plotted for $\epsilon$ across $[0,\frac12)$. Initial conditions were selected using random sampling of phase space, in a way to render, if not all, then the most essential attractor features, with an emphasis on detecting asymmetry. 
\begin{figure}[ht]
\begin{center}
\includegraphics*[width=28mm]{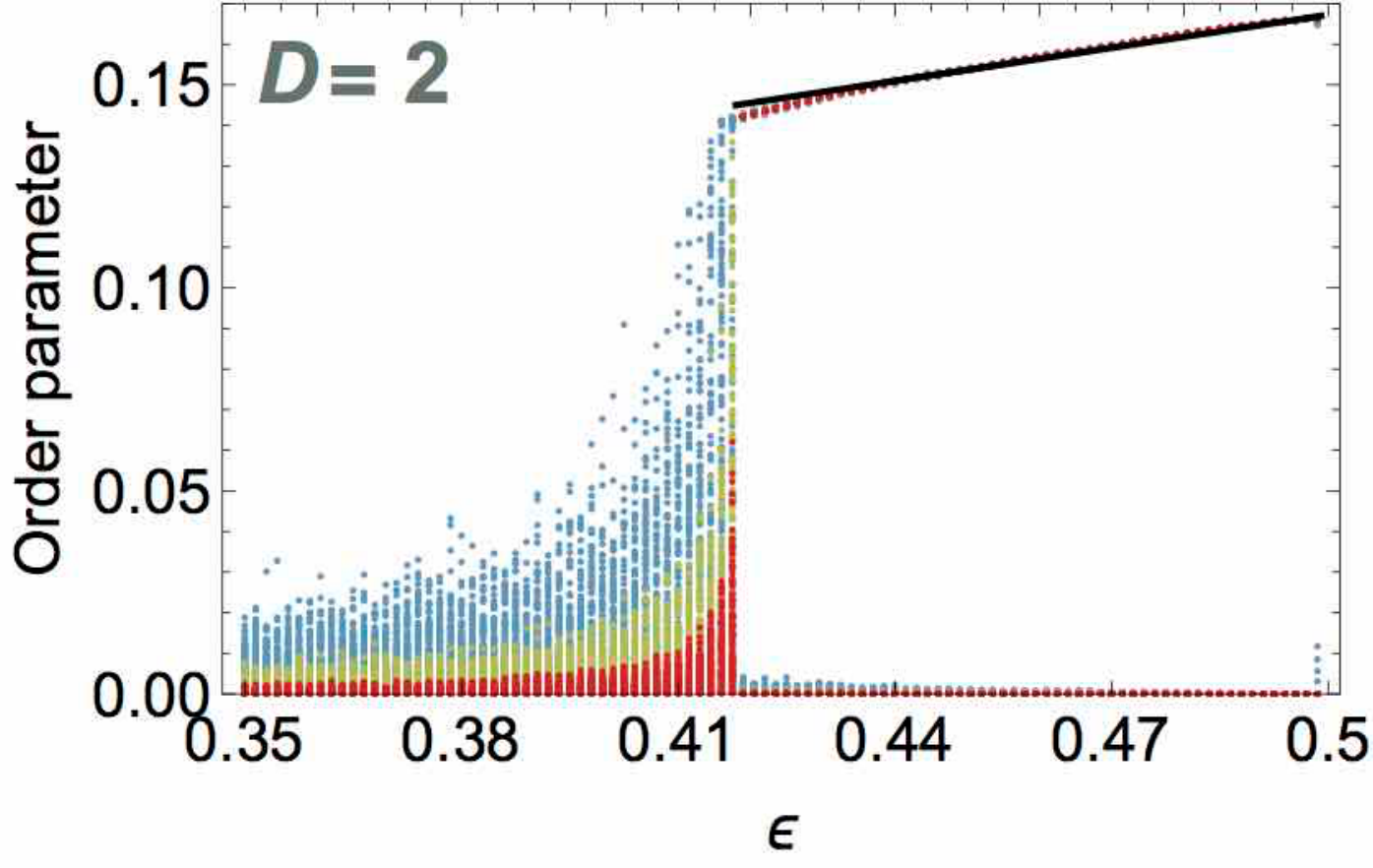}
\includegraphics*[width=28mm]{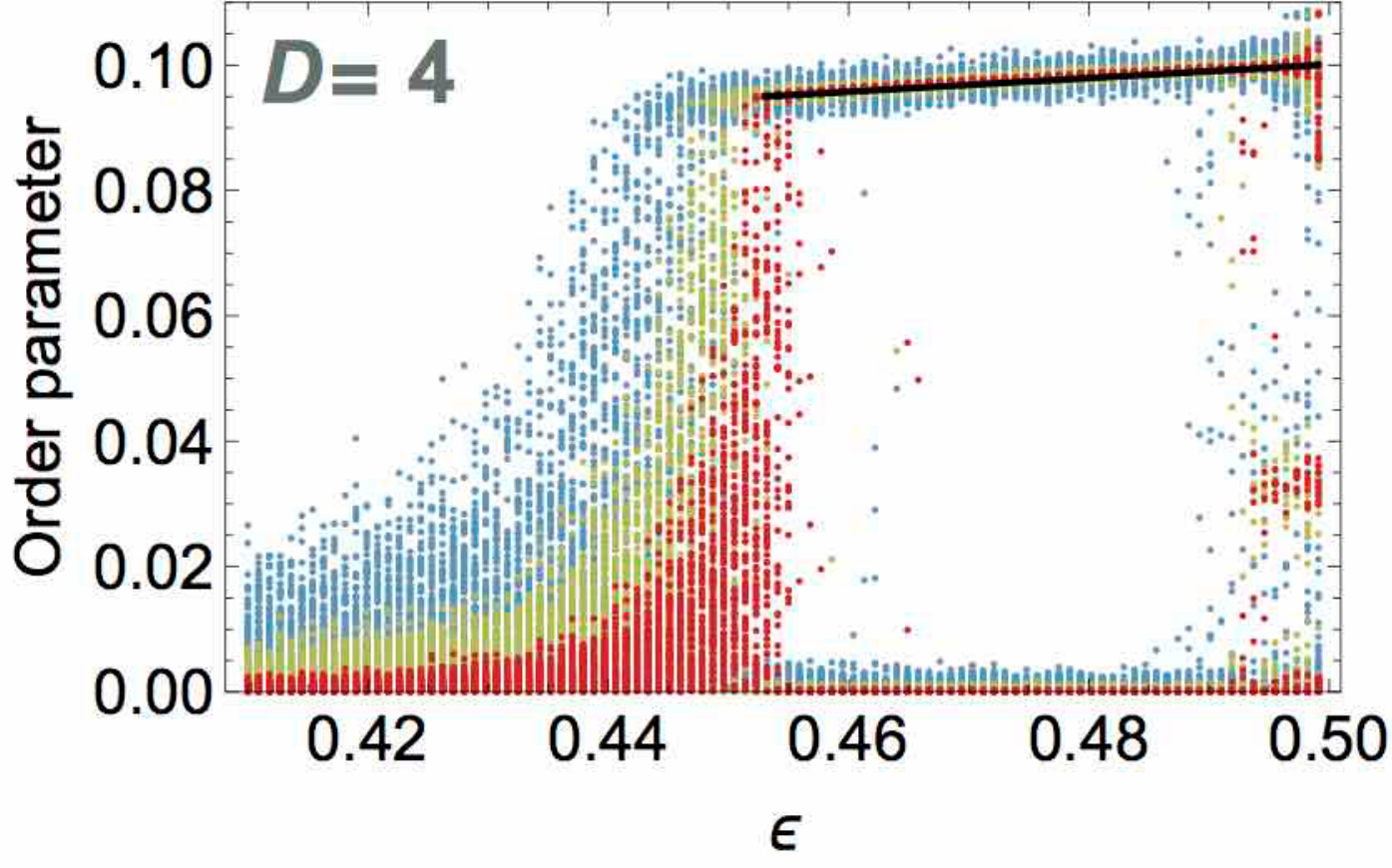}
\includegraphics*[width=28mm]{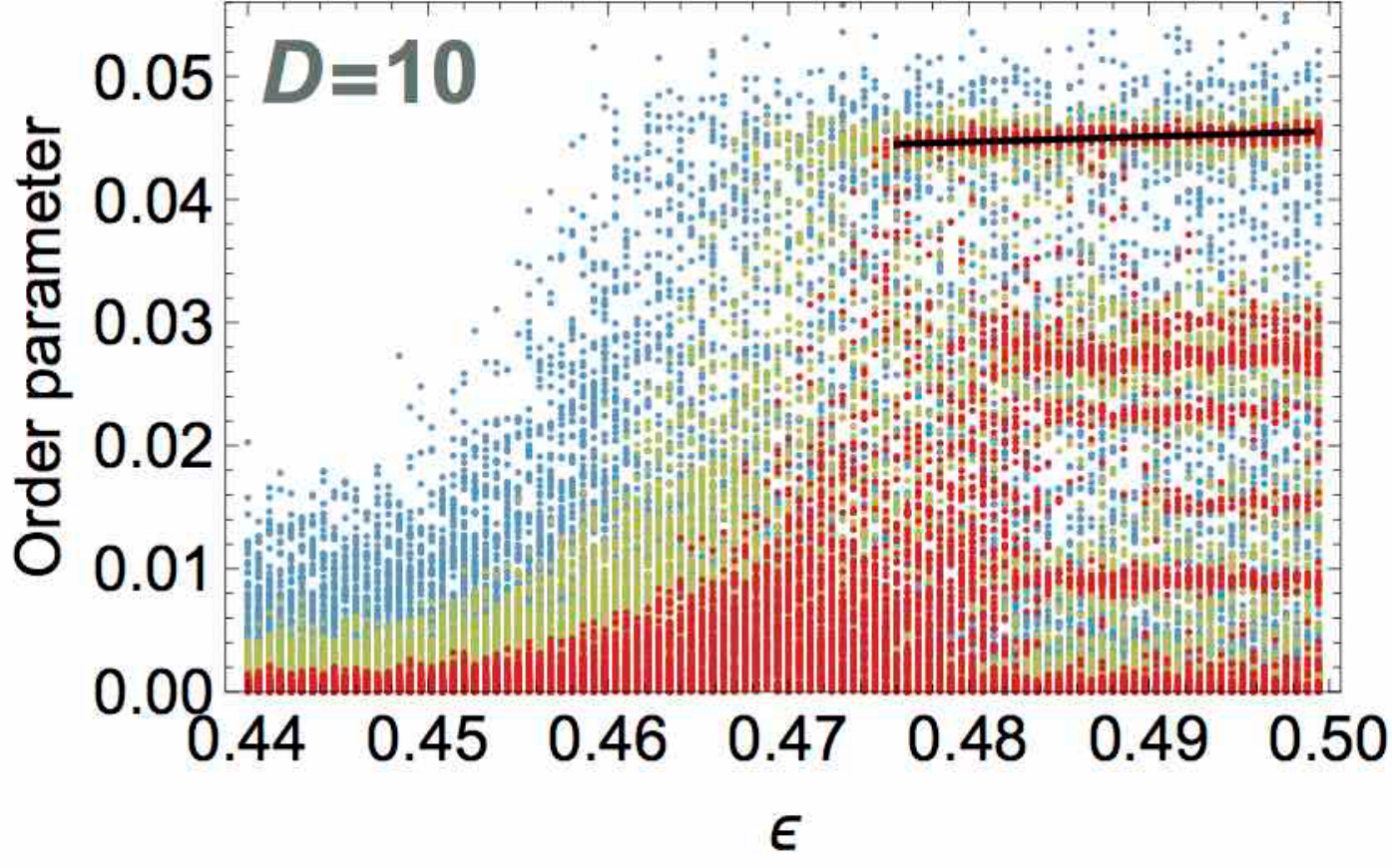}
\includegraphics*[width=28mm]{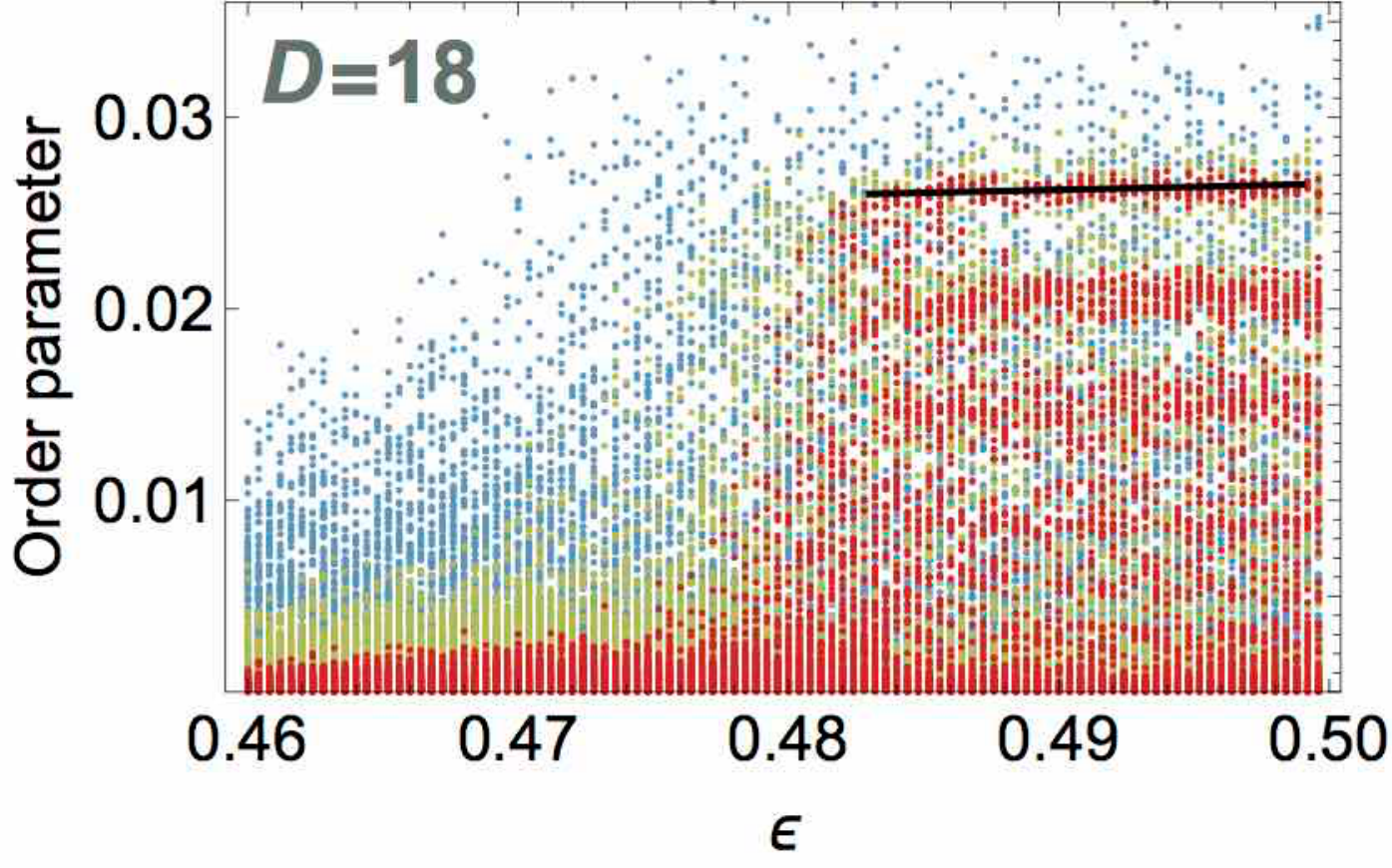}

\vspace{0.2cm}
\includegraphics*[width=28mm]{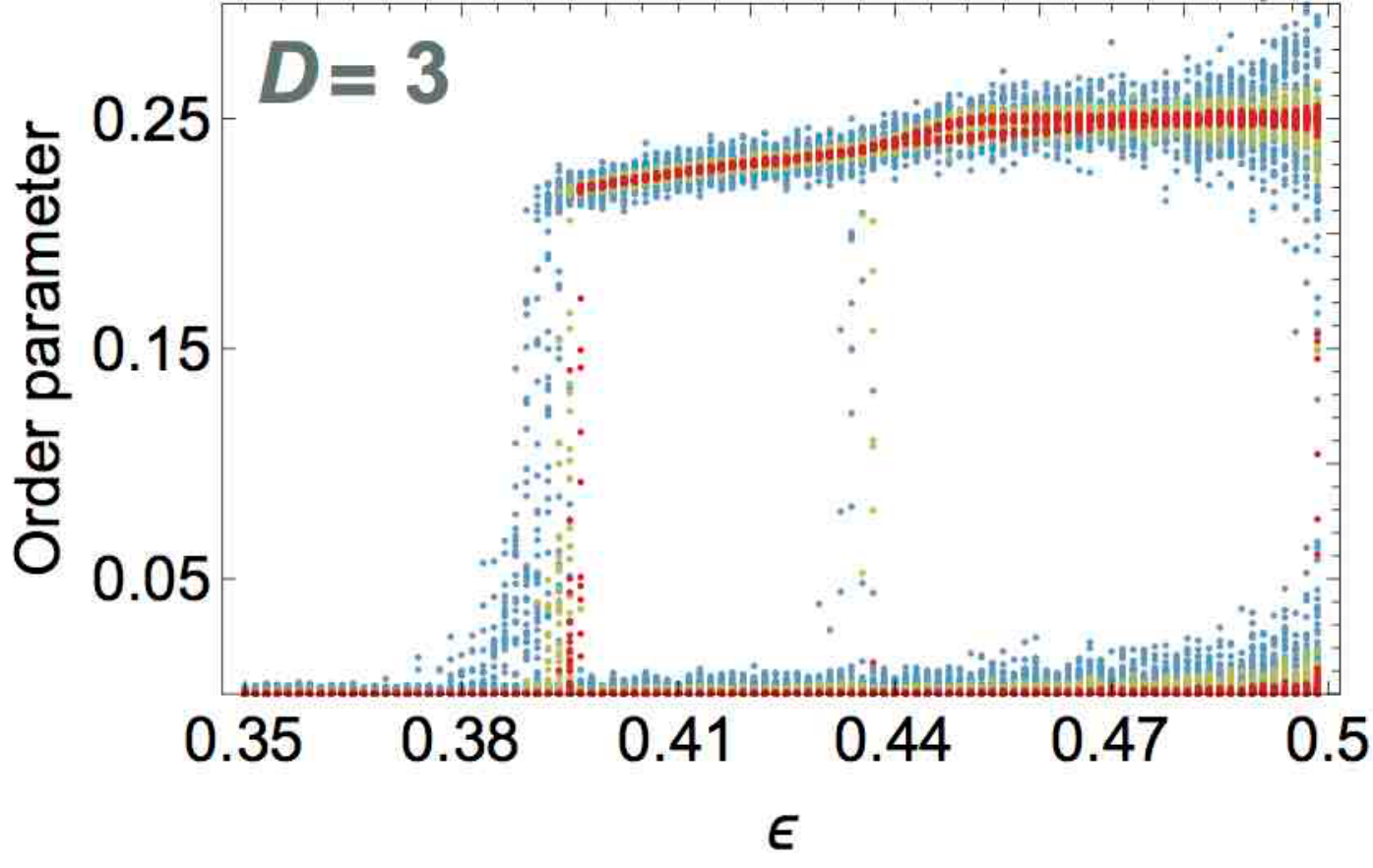}
\includegraphics*[width=28mm]{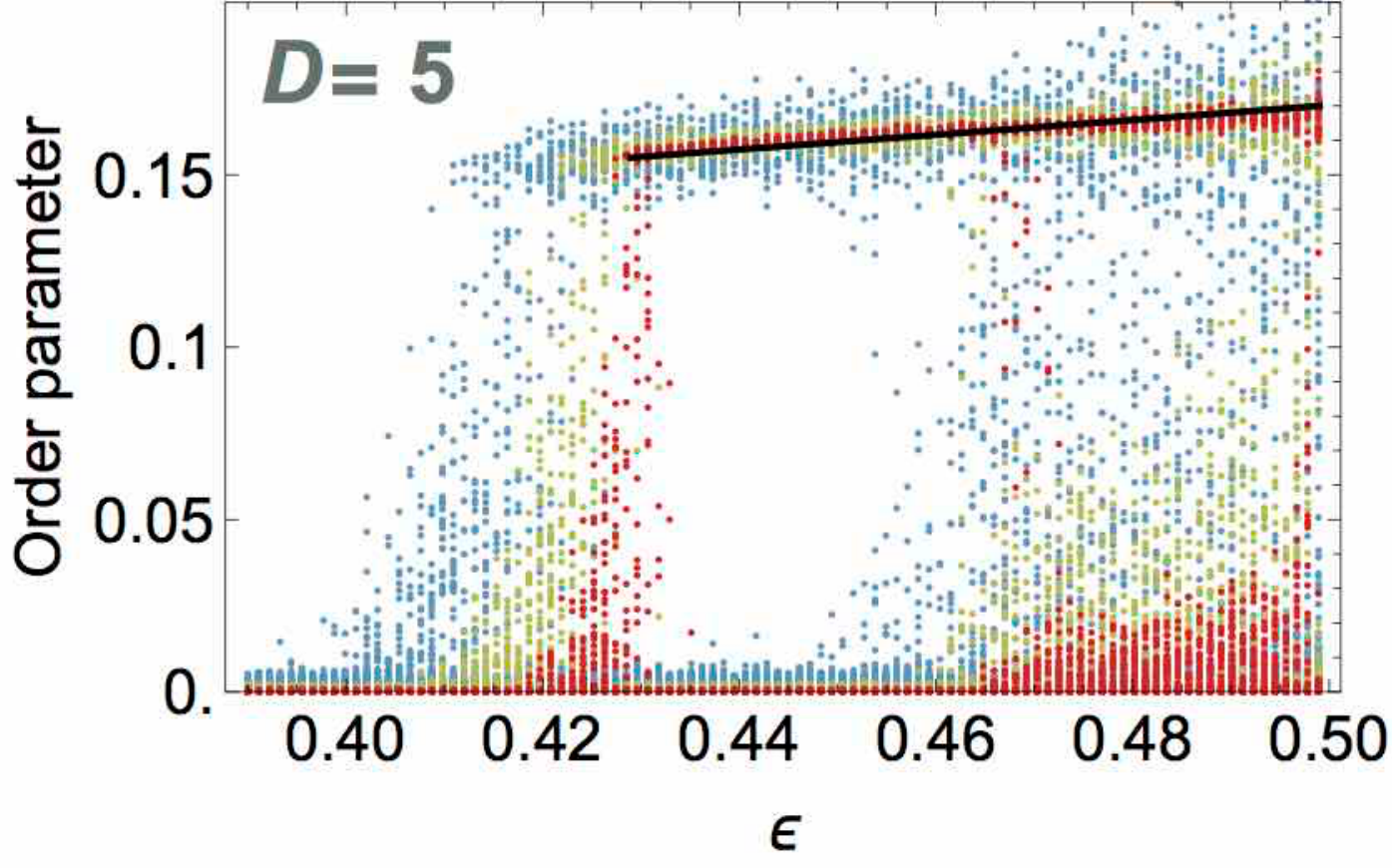}
\includegraphics*[width=28mm]{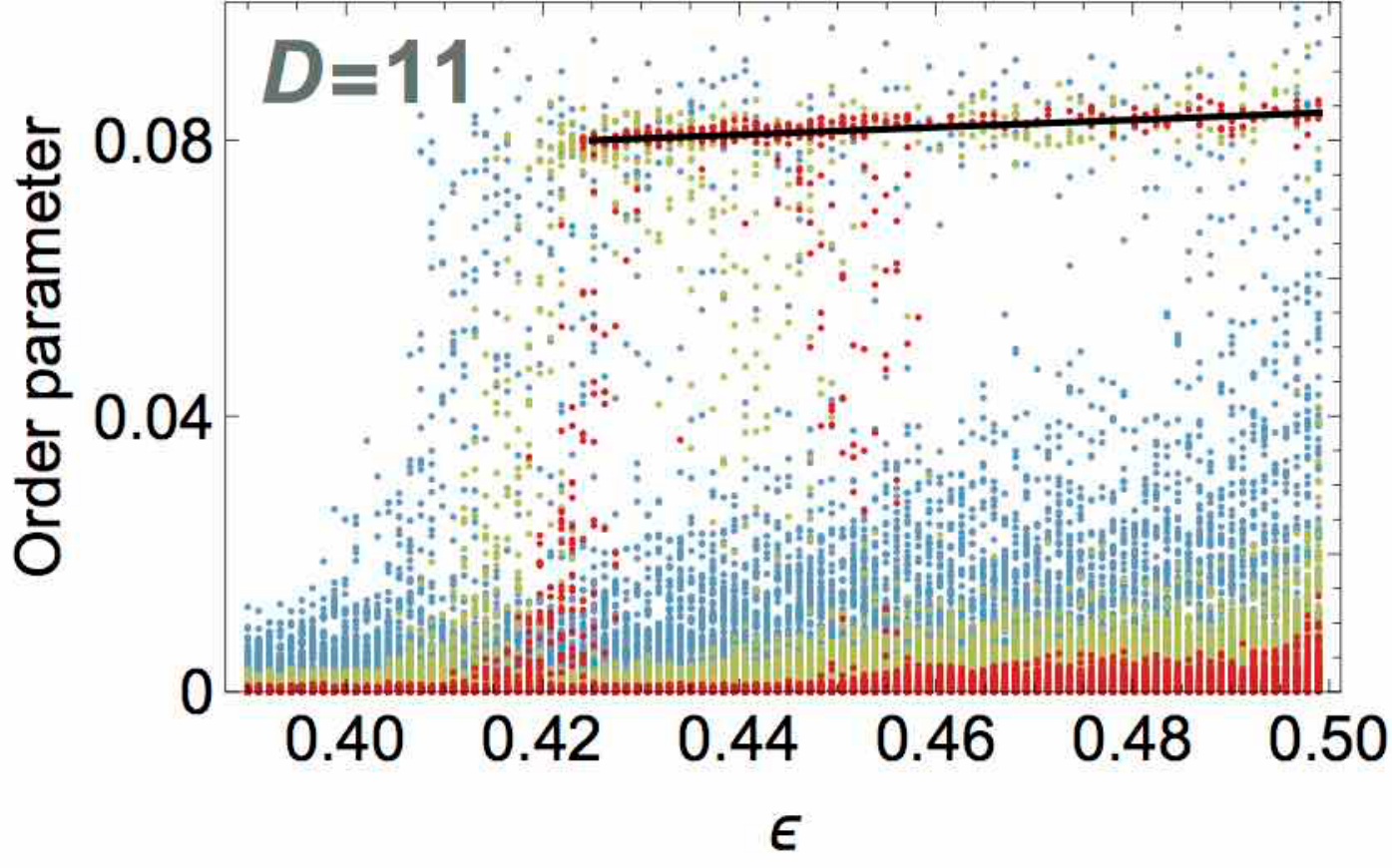}
\includegraphics*[width=28mm]{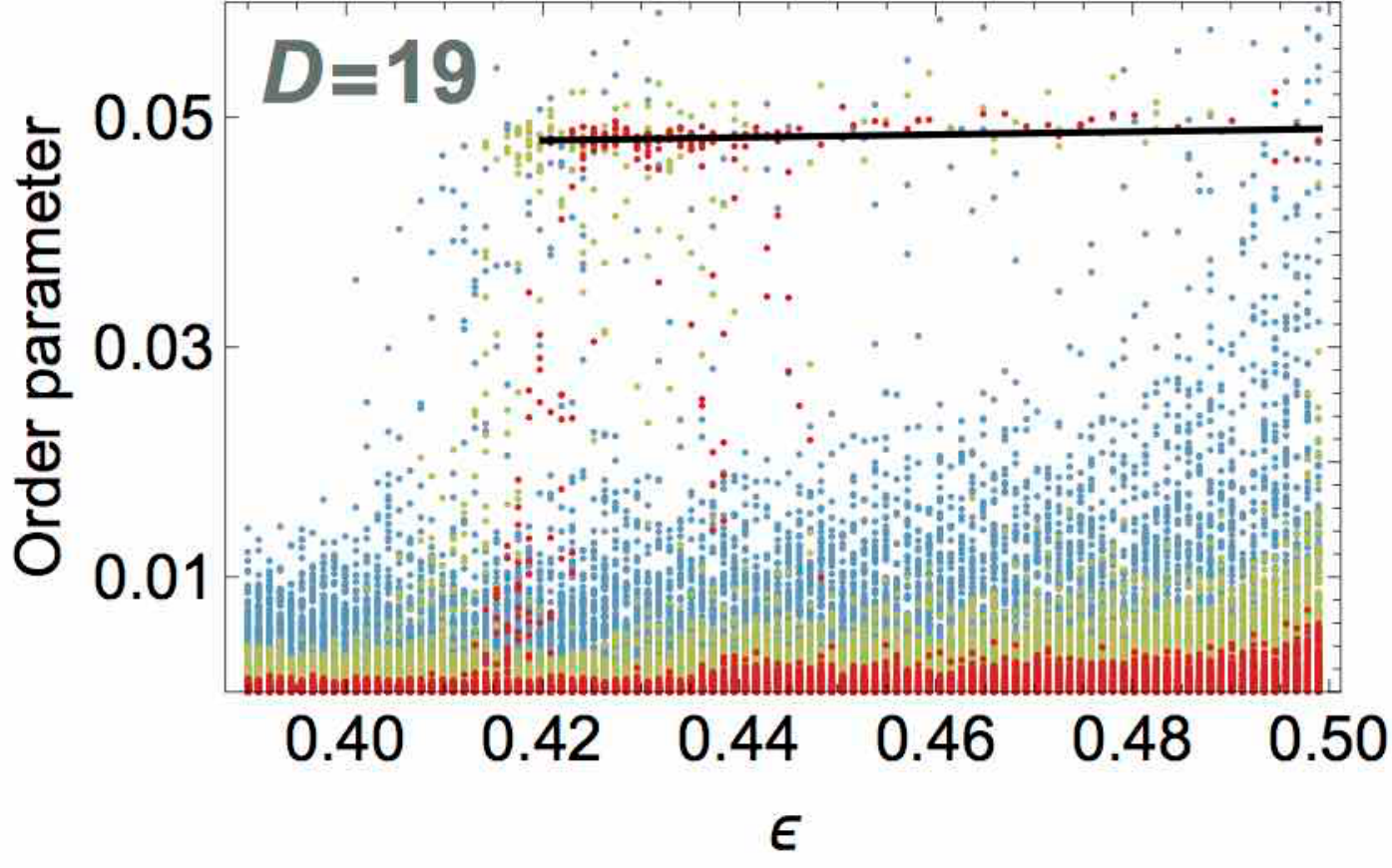}

\vspace{0.2cm}
\includegraphics*[height=17mm]{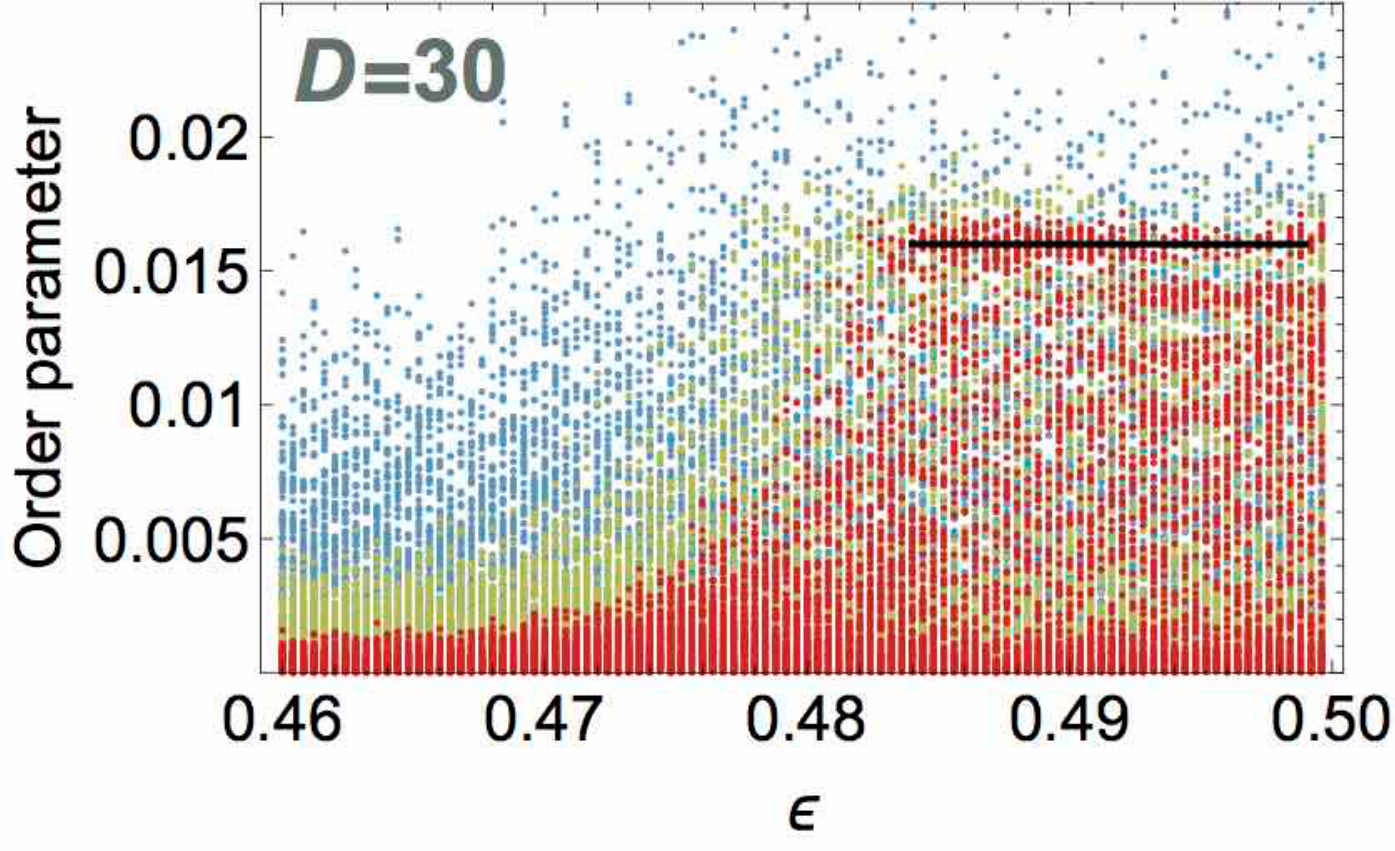}
\includegraphics*[height=17mm]{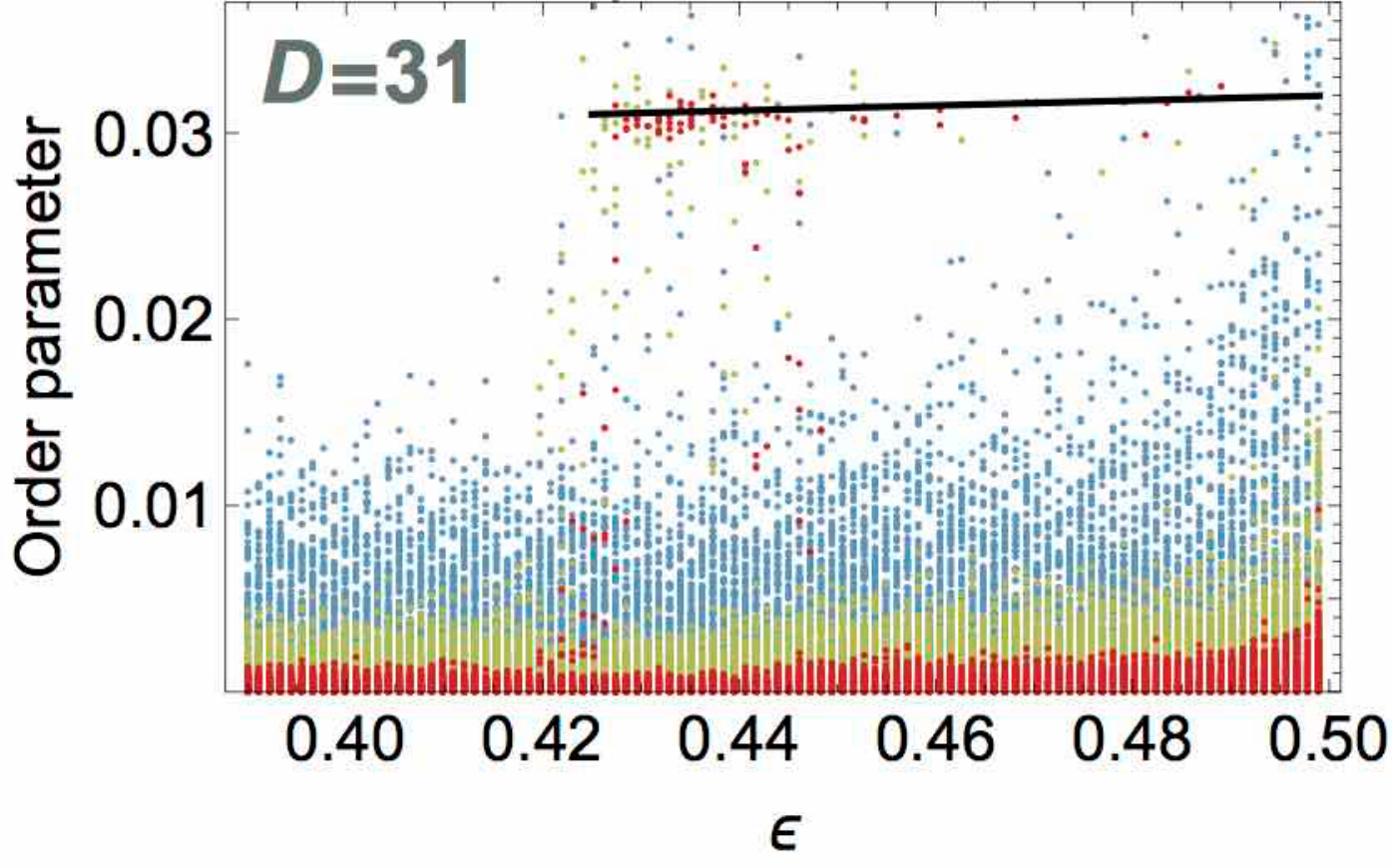}
\includegraphics*[height=17mm]{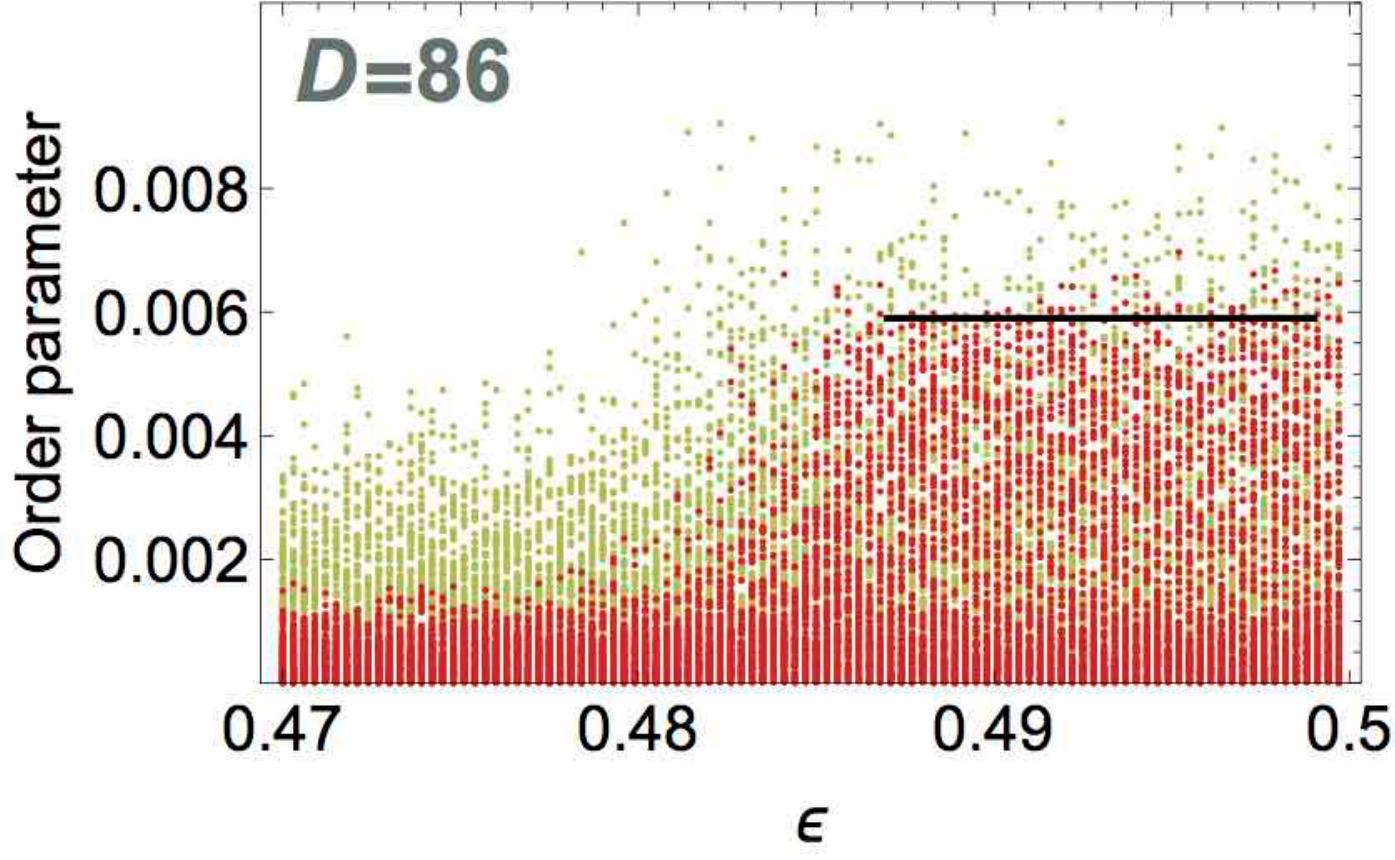}
\includegraphics*[height=17mm]{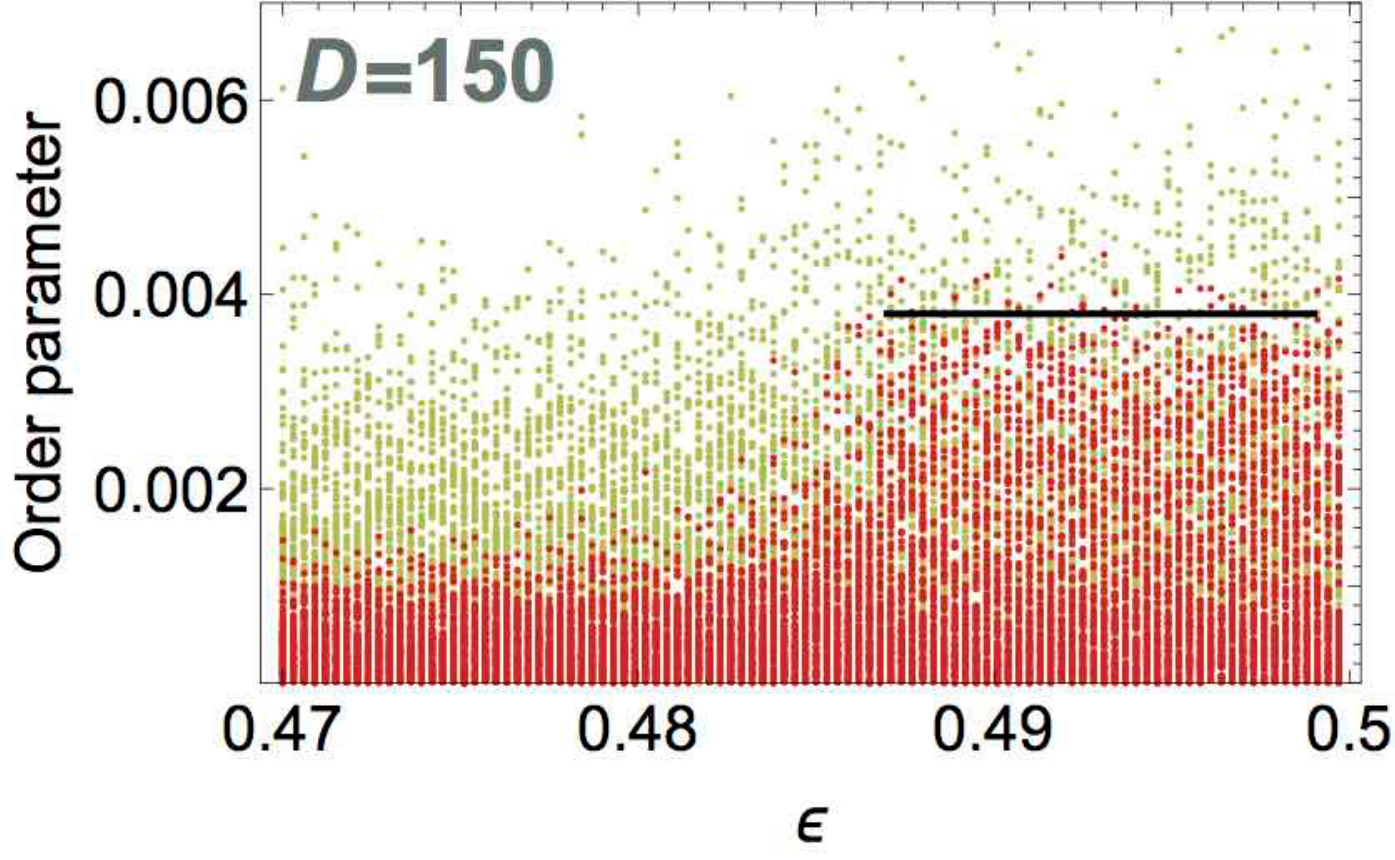}

\vspace{0.2cm}
\includegraphics*[height=17mm]{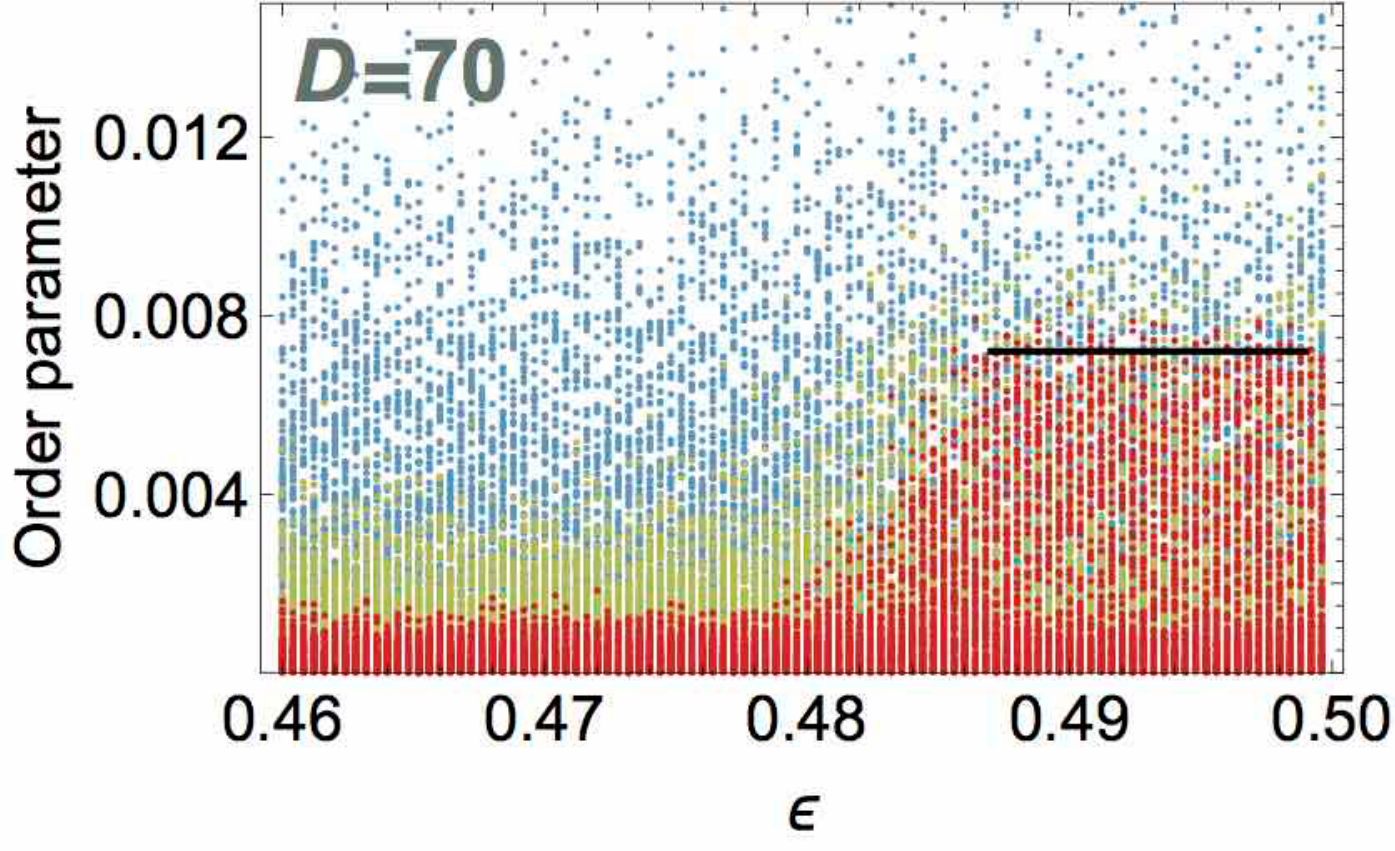}
\includegraphics*[height=17mm]{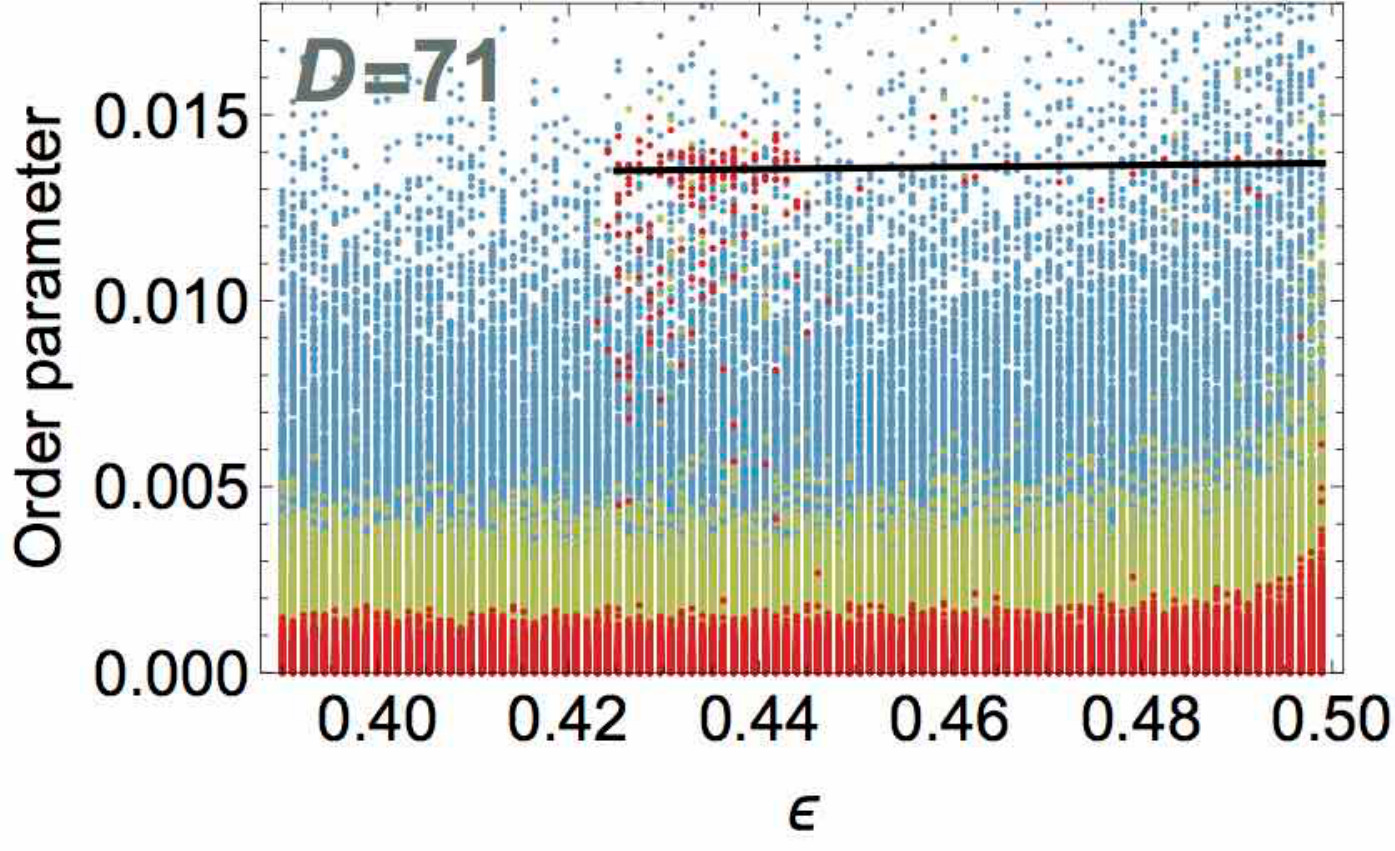}
\includegraphics*[height=17mm]{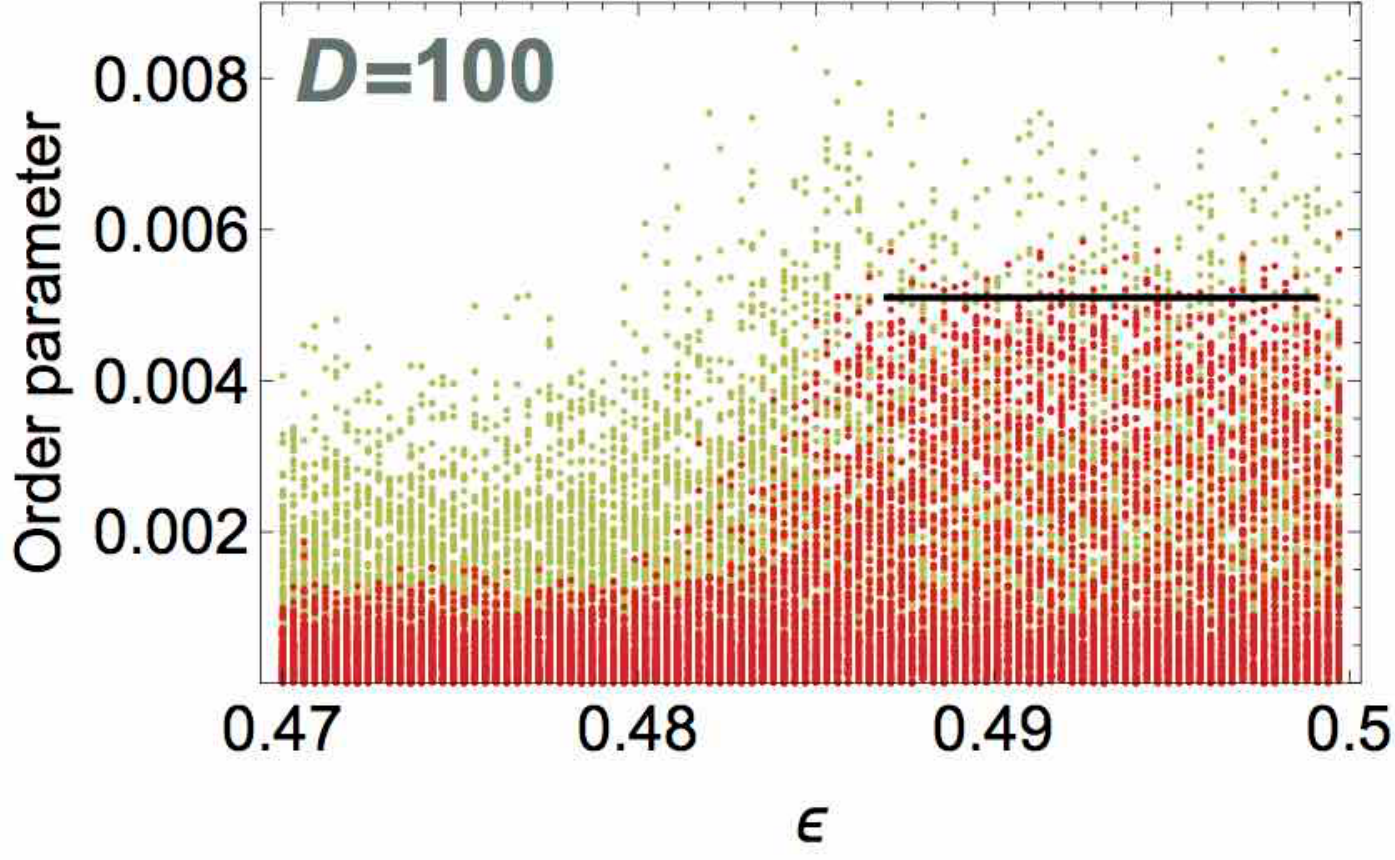}
\includegraphics*[height=17mm]{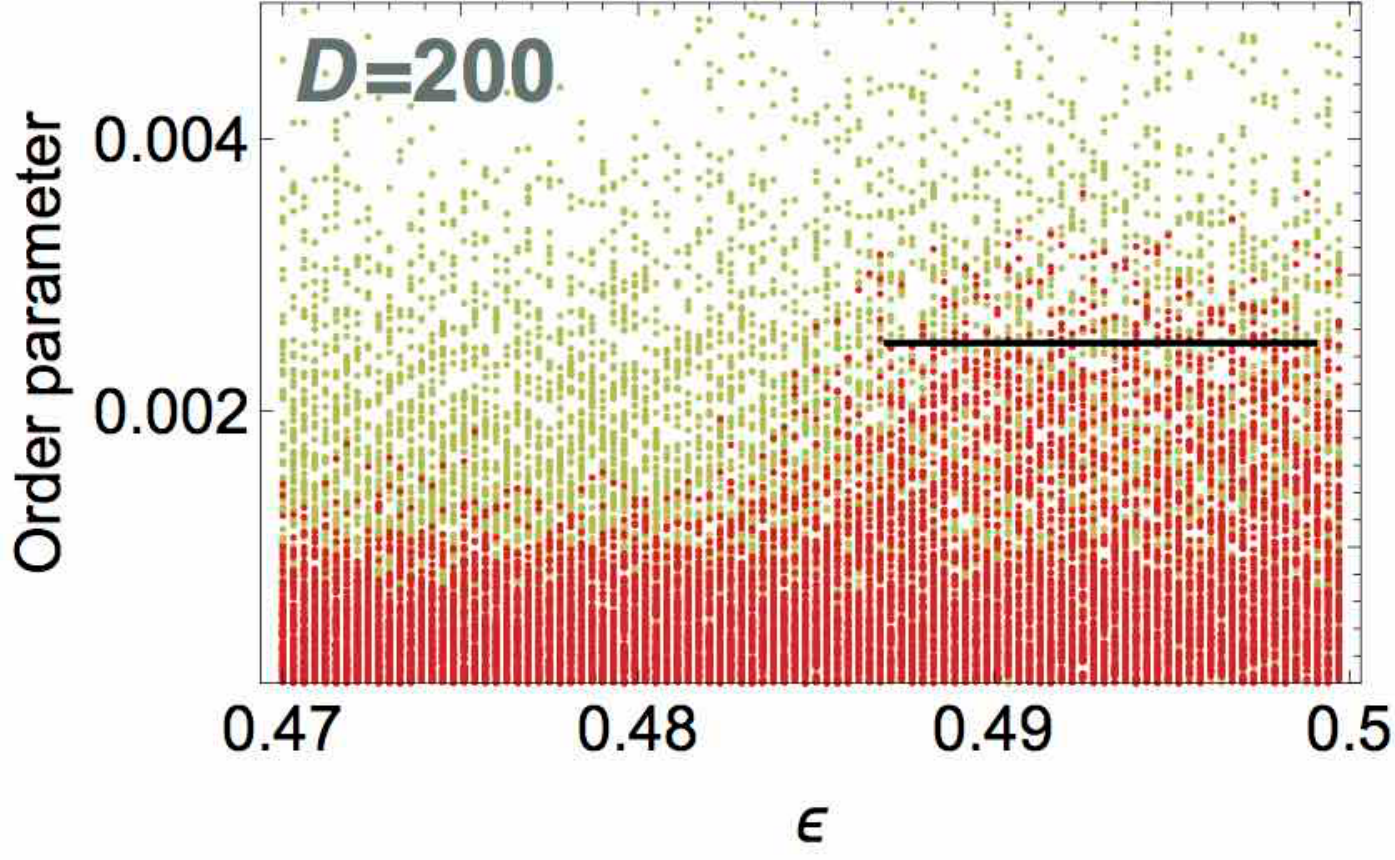}
\end{center}
\caption{Empirical estimates of the sign-flip asymmetry observable. Superimposed plots of the order parameter $\left|\frac1{T}{\displaystyle\sum_{t=1}^{T}}x^t_{\lceil\frac{D}2\rceil}-\tfrac12\right|$ for the projected points $x^t= \pi_{(0,1)^D}\circ G_{D,\epsilon}^t(x)$ in $(0,1)^D$ of orbits started from random initial conditions $x$ drawn from the uniform distribution. In each picture, 100 values, corresponding to 100 initial conditions, are plotted for each value of $\epsilon$ and each color; first blue ($T=10^4$), then green ($T=10^5$) and finally red ($T=10^6$). The process is repeated for 100 values of $\epsilon$. For $D\geq 70$, fluctuations of the values for $T=10^4$ are too large to assert symmetry breaking. For the sake of clarity, these estimates are not reported on the pictures for $D\geq 86$. Black segments: linear interpolation of maximum order parameter in the non-ergodic regime.}
\label{ORDERPAR}
\end{figure}

Although the bifurcation scenarios differ for the two cases, the pictures reveal that when increasing $\epsilon$ from 0, ergodicity persists for $\epsilon$ up to some $\epsilon_D$ and then fails beyond that threshold. In the $D=2$ case, the transitive and fully symmetric attractor continuously splits at $\epsilon=\epsilon_2$ into 6 disjoint and asymmetric invariant pieces. Each emerging piece breaks all map symmetries except one (Appendix \ref{SymBreakG2}). In the $D=3$ case, a fully symmetric invariant set exists throughout the expanding domain. In addition, 6 asymmetric invariant components discontinuously appear at $\epsilon=\epsilon_3$, away from the symmetric set, and persist from thereon as $\epsilon$ continues to increase. Then, at $\epsilon\simeq 0.437$, this group of partly asymmetric orbits is augmented in a similarly discontinuous way by an additional analogously persisting group, composed of 8 asymmetric orbits (see the involved symmetries in Appendix \ref{SymBreakG3}). Despite that the phenomenology differs in the two cases, the asymmetric components always appear to be disjoint from their image under the sign-flip, {\sl ie.}\ the $\Z_2$-symmetry generated by $-\text{Id}_{\T^D}$ is systematically broken.

The analytic proofs of ergodicity breaking in \cite{F14,S18,SB16} established the existence of so-called InAsUP (see Definition 1 below) for all $\epsilon$ larger than thresholds that are remarkably close to the $\epsilon_D$ above. InAsUP were guessed using trajectory renderings as above.

For large(r) $D$, trajectory renderings are obviously more involved and certainly not so simply useful to detect ergodicity/symmetry breaking. Following standard diagnostics in statistical physics, one can instead use order parameter (OP) empirical estimates. An empirical OP consists of unsigned averages over consecutive iterates of an asymmetry-related observable, which is designed to suggest the existence of asymmetry ergodic components when positive. In order to establish failure of sign-flip symmetry, we use the "central" coordinate $x_{\lceil\frac{D}2\rceil}$ as an observable (Fig.\ \ref{ORDERPAR}); other estimates based on different quantifiers {\sl e.g.}\ any single coordinate, two/all coordinate mean values, etc., all yield similar plots with identical bifurcation values (data not shown). Finite-time effects are accounted for by superimposing results from averages over increasing numbers of iterates. Similarly, dependence on the initial condition is evaluated using multiple runs based upon randomly drawn inputs. 

In agreement with ergodicity, the OP in Fig.\ \ref{ORDERPAR} vanishes at small coupling, for every $D$. However, as soon as $\epsilon$ exceeds some $\epsilon_D$, this quantity takes on positive values for a positive fraction of initial conditions. This was observed for all  investigated values of the dimension, from $D=2$ up to $D=200$. For $D=2$ and $3$, the data are consistent with the phase space plots of Fig.\ \ref{PHASPA}; the emergence of positive OP coincides with the appearance of asymmetric ergodic components.\footnote{Surprisingly, for $D=3$, the additional asymmetric group emerging at $\epsilon\simeq 0.437>\epsilon_3$ in Fig.\ \ref{PHASPA} -- signs of the corresponding transition are barely visible on Fig.\ \ref{ORDERPAR} --  shows OP values that are very close, if not identical, to the primary asymmetric group.} 

In addition to clear evidence of ergodicity breaking, Fig.\ \ref{ORDERPAR} reveals various interesting phenomenological features of the maps $G_{D,\epsilon}$ dynamics. At first, the bifurcation diagrams show characteristics that are specific to the parity of $D$. For $D$ odd, the bifurcation values $\epsilon_D\sim 0.42$ appear to be almost insensitive to $D>3$, and OP estimates are localized around a single ($\epsilon$-dependent) value.  Moreover, the fraction of initial conditions that yield non-zero estimates decreases with $D$, making it more difficult to collect marked evidence of ergodicity breaking. For $D$ even, $\epsilon_D$ increases with $D$ and seems to approach a limit value $\epsilon_{\ast}<\frac12$.
OP estimates appear to be uniformly distributed between 0 and the ($\epsilon$-dependent) maximal value, making it difficult to discriminate limit values from finite size fluctuations. 

\begin{figure}[ht]
\centering
\includegraphics*[height=30mm]{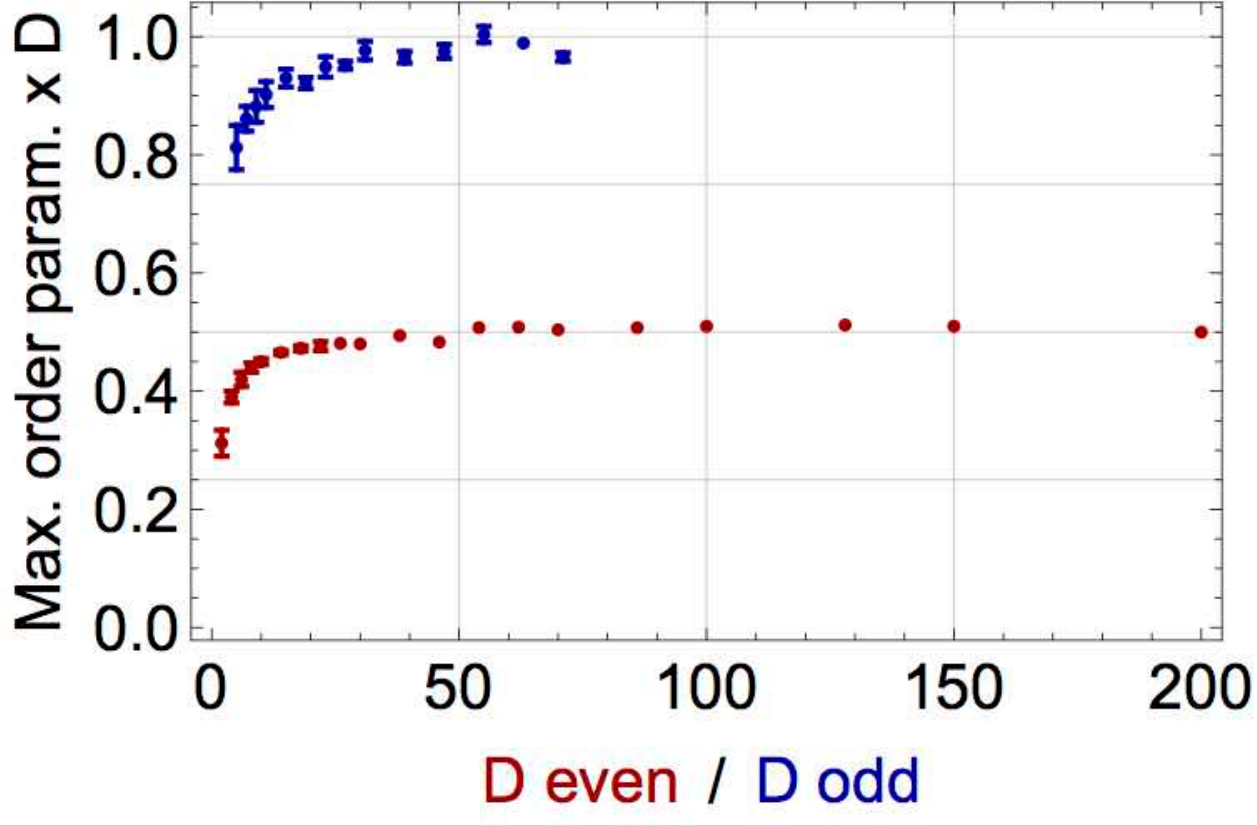}
\caption{Maximal order parameter range$\times D$ vs.\ $D$, respectively for even and odd values of $D$.}
\label{MAXOP}
\end{figure}

Furthermore, 
maximal OP values for $\epsilon>\epsilon_D$ show a linear $\epsilon$-dependence of tiny slope, which eventually vanishes for large $D$. Also, these maxima decrease as $D$ increases, and asymptotically behave as $\frac1{2D}$ for $D$ even (resp.\ $\frac1{D}$ for $D$ odd), see 
Fig.\ \ref{MAXOP}.
Accordingly, to unambiguously distinguish asymmetry from short time fluctuations requires longer averages when $D$ increases.\footnote{In particular, $T=10^4$ averages do not suffice to identify the emergence of positive OP for dimensions $D\geq 70$.} This issue, when combined with linear increase of the dimension of the variables in the iterations, substantially increases the computation time required for conclusive evidence. For instance, to obtain the $100\times 100$ averages over $T=10^6$ iterates in each of the pictures for $D\geq 70$ required running times of $\sim 100\text{h}$ on a $2.4$ GHz multi-processor computer (compared to $\sim 1\text{min}$ for $D=2$). Therefore, to show ergodicity breaking for dimensions that are commensurate with realistic physical systems appears to be a considerable challenge. 

\section{Computer proof of ergodicity breaking}
The previous section evidences of emergence of asymmetric Lebesgue ergodic components 
call for rigorous confirmation, in particular to exclude transient effects and other computer round-off shortcomings that may impact numerical simulations and finite-time estimates. A computer-based rigorous proof of ergodicity breaking is presented in this section.

For piecewise affine and expanding maps such as $G_{D,\epsilon}$, any forward invariant finite union of (convex) polytopes must support an absolutely continuous invariant measure \cite{T01}. Therefore, in order to prove existence of asymmetric Lebesgue ergodic components,
it suffices to obtain such unions of polytopes that are disjoint from their image under $-\text{\rm Id}_{\T^{D}}$.\footnote{Throughout this section, a set $S$ is said to be {\bf asymmetric} iff $S\cap -\text{\rm Id}_{\T^{D}}(S)=\emptyset$.} Accordingly, we shall rely on the following notion. 
\begin{Def}
Let $D,\epsilon$ and a finite union of (non-empty) polytopes $P\subset \T^D$ be given. Then $P$ is said to be an {\bf InAsUP} (acronym for Invariant Asymmetric Union of Polytopes) if it satisfies the following conditions
\[
G_{D,\epsilon}(P)\subset P\quad\text{and}\quad P\cap -\text{\rm Id}_{\T^{D}}(P)=\emptyset.
\]
\end{Def}
As already mentioned, analytic proofs of existence of InAsUP have been established for $D=2,3$. However, the proofs rely on observations of trajectories and would be hardy generalisable to large values of $D$. Instead, a fully computational approach is proposed below, which consists of an algorithm designed to generate InAsUP for arbitrary $D$. 

\subsection{Principles of the algorithm and exact computer results}
In few words, the algorithm simply consists in applying repeated iterations of $G_{D,\epsilon}$ to an asymmetric initial polytope.\footnote{See subsection \ref{S-ALGO} below for more details and a pseudo-code is given in Appendix \ref{A-ALGO}).} It 
terminates when the resulting union set becomes $G_{D,\epsilon}$-invariant, or prematurely stops if the set under construction happens to intersect it image under $-\text{\rm Id}_{\T^{D}}$. 

As analytic proofs did, the algorithm optimizes its input by using dynamical information from simulations. Initial polytopes are chosen among cylinder sets that are given by symbolic codes associated with empirical trajectories of positive OP (subsection \ref{S-ADAPT}). 

That initial polytopes are cylinder sets is crucial when $\epsilon\in\Q$ because all algorithmic calculations then involve rational numbers only (subsection \ref{S-POLDYN}). Using exact computer arithmetics on such numbers\footnote{In particular, we use the GNU arithmetic library GMP \cite{GMP}.}, this implies that when the construction completes, the resulting asymmetric invariant set must be a genuine InAsUP. In other words, when the construction completes, the computer provides a rigorous proof of ergodicity breaking for the pair $(D,\epsilon)$ under consideration. 

To employ exact arithmetic is an important feature of the InAsUP algorithmic construction. Indeed, analytic proofs for $D=2,3$ have revealed that some polytope facets are exactly mapped onto other ones, even when $\epsilon$ is an irrational number. Hence, some analytic cancellations must take place in the construction, which the algorithm would have to carefully monitor, if it dealt instead with floating-point arithmetic. In short terms, in using floating-point arithmetic, control of round-off errors does not suffice to rigorously assert invariance; hence the choice of exact arithmetic here.


Results on exact computer InAsUP constructions are summarized in the following formal statement. 
\begin{Claim}
For all values of the pair $(D,\epsilon)$ in Table \ref{ALGOTAB}, the map $G_{D,\epsilon}$ has an InAsUP.
\label{COMPCLAIM}
\end{Claim}
\begin{table}
\caption{Data summary of InAsUP's exact numerical construction}
\begin{center}
\begin{tabular}{|c|c|c|c|c|c|c|}
\hline
\multicolumn{7}{|c|}{Main results}\\
\hline
D&$N_D$&$\epsilon$&$|$Cyl$|$&Succ.\ ratio&$\#$ InAsUP&CPU time\\
\hline
2&3&0.44&5&24/24&22 (43)&1.5ms (3.5ms)\\
\hline
3&13&0.4&9&11584/11706&1250 (9100)&0.6s (13.9s)\\
\hline
4&75&0.47&11&355/373$^\ast$&3.6 (6.2)$\times10^5$&1.4h (5.2h)\\
\hline
5&541&0.44&7&3/3$^\ast$&6 (10)$\times 10^6$&120h (200h)\\
\hline
\end{tabular}
\end{center}
\noindent
\begin{tabular}{p{12cm}}
{\sl Legend:}\\
$N_D=$ cardinality of atomic partition ({\sl ie.}\ numb. of atoms in $S_D$)\\
$|$Cyl$|$ = length $\ell+1$ of words $a_0\cdots a_\ell$ that define initial cylinders\\
Succ.\ ratio = $\frac{\text{numb.\ of cylinders for which the construction succeeded}}{\text{numb.\ of cylinders for which the construction ran}}$. The construction ran for each cylinder generated by some empirical trajectory with positive OP ($^\ast$ = the construction ran for less cylinders, for the sake of computation time.)\\
$\#$ InAsUP = typical (resp.\ maximal) number of polytopes in InAsUP\\
$\#$ CPU time = typical (resp.\ maximal) computation time for InAsUP completion (CPU frequency = $2.2$ GHz)
\end{tabular}
\label{ALGOTAB}
\begin{center}
\begin{tabular}{|c|c|c|c|}
\hline
\multicolumn{4}{|c|}{Additional InAsUP success ratios}\\
\hline
D&$\epsilon$&$|$Cyl$|$&Succ. ratio\\
\hline
2&0.44&2&2/6\\
&&3&8/10\\
\hline
3&0.4&2&12/61\\ 
&&4&398/585\\
\hline
4&0.47&5&0/29$^\ast$\\
&&6&7/43$^\ast$\\
\hline
5&0.44&4&3/27$^\ast$\\
&&5&1/6$^\ast$\\
\hline
\end{tabular}
\end{center}
\end{table}
To our best knowledge, these results provide the first proof of ergodicity breaking of $G_{D,\epsilon}$ for $D\geq 4$,  and hence of the coupled maps $F_{N,\epsilon}$ for $N\geq 5$. For $D=2,3$, the results confirm the previous analytic proof conclusions. For $D\leq 4$, InAsUP have also been obtained for other values of $\epsilon$ (data not shown). As Fig.\ \ref{ORDERPAR} suggests, we expect InAsUP to exist for every $\epsilon>\epsilon_D$. 

In addition, Table \ref{ALGOTAB} provides success ratio statistics against the length of the initial cylinder. 
Clearly, the larger the length is, the higher a success ratio results. This suggests that it suffices to choose a sufficiently small neighborhood of any point of some asymmetric trajectory to generate an InAsUP. Not only ergodicity breaking is not a spurious numerical illusion, but transient behaviors and round-off errors in simulations (Fig.\ \ref{ORDERPAR}) do not impact this phenomenon.

For completeness, Table \ref{ALGOTAB} also provides statistics about InAsUP cardinality and related CPU computation times ({\sl viz.}\ typical value and maximum).\footnote{As explained in the end of subsection \ref{S-ADAPT}, polytopes in InAsUP may overlap. InAsUP cardinality thus makes no obvious sense other than justifying the measured CPU times.} These quantities show substantial variations upon initial cylinder. However, their variation ranges barely vary with the cylinder length. Furthermore, Table \ref{ALGOTAB} reveals rapid growth in these statistics - which accompanies an exponential growth of the atomic partition cardinality - as the dimension $D$ increases. Therefore, even though the construction could in principle operate for any $D$, these exploding features, which imply similar demand of computational resources, may prevent the construction to terminate for large $D$.\footnote{In particular, for $D=6$ ($N_6=4683$), no construction has terminated, due to insufficient available RAM and CPU time.}  

\subsection{Details of the algorithm for InAsUP construction and its numerical implementation}
A copy of the source files of the InAsUP construction code (in C) is available online \cite{F19} (and, again, see Appendix \ref{A-ALGO} for a pseudo-code). Most of the procedures in the algorithm may appear to be standard. However, to manipulate and to apply geometric and topological operations on polytopes in arbitrary dimension (such as testing inclusion, testing intersection, computing intersection set, etc), seem not so common.  Thanks to the isotropic nature of the $G_{D,\epsilon}$, a convenient, dimension-independent, approach to polytope manipulation and their operations can be developed in our setting. The purpose of this section is to provide insights into these specific procedures and their numerical implementation. 

For convenience, we regard $G_{D,\epsilon}$ as a mapping from $S_D=(-\tfrac12,\tfrac12)^D$ into $\T^D$ (to be combined with the canonical projection from $\T^D$ to $S_D$), and we denote by 
\[
\bigcup_{a=1}^{N_D}A_a\ \text{mod}\ 0,
\]
the partition of $S_D$ into polytopes $A_a$ - called {\bf atoms} - on which the mapping is affine (or, equivalently, on which the vector function $B_D$ is constant). Important for future purposes, every atom is a polytope whose facets are included in discontinuity planes (or in parallel planes). Given the expression of $B_D$, every discontinuity plane is characterized by the condition $h(\sum_{k=i}^jx_k)\in\tfrac12+\Z$ for some $i<j$.

\subsubsection{Algorithmic procedure}\label{S-ALGO}
The InAsUP construction algorithm can be sketched as follows:
\begin{itemize}
\item[$\bullet$] Choose an initial polytope $P^0\subset\T^D$ such that $P^0\cap -\text{\rm Id}_{\T^{D}} (P^0)=\emptyset$.  
\item[$\bullet$] Compute subsequent iterates $P^{t+1}=G_{D,\epsilon}(P^t)$ for $t\in\N$ -- which consist of unions of polytopes\footnote{More precisely, $P^{t+1}$ consists of a single polytope as long as ($t$ is such that) $P^t\subset A_a$ for a single $a$. However, as soon as we have $P^t\cap A_{a_i}\neq \emptyset$ for several $i$, then
\[
G_{D,\epsilon}(P^t)=\bigcup_i G_{D,\epsilon}|_{A_{a_i}}(P^t\cap A_{a_i})\ \text{mod}\ 0
\]
{\sl viz.} $G_{D,\epsilon}(P^t)$ becomes the union of several polytopes (mod 0), and therefore so must be all subsequent images $P^{t'}$ for $t'>t$.} -- until either $P^{t+1}\subset \bigcup_{k=0}^{t}P^k\ \text{mod}\ 0$, or some polytope in $-\text{\rm Id}_{\T^{D}} (P^{t+1})$ intersects $\bigcup_{k=0}^{t+1}P^k$.
\end{itemize}
As explained before, the purpose of the second terminating condition is to interrupt the construction when guarantee fails that the constructed sets will be asymmetric. Otherwise, when the algorithm terminates in the first instance, then both resulting $\bigcup_{k=0}^{t}P^k$ and $-\text{\rm Id}_{\T^{D}} (\bigcup_{k=0}^{t}P^k)$ must be disjoint InAsUPs, as desired. 

The map $G_{D,\epsilon}$ does not show any basic property, such as Markov partition \cite{KH95}, that would ensure the algorithm to always terminate in finite time. However, this is the case of all results in Table \ref{ALGOTAB}, {\sl viz.}\ when the construction does not succeed, intersection with symmetric image is always the cause. 


\subsubsection{Adapted polytopes and their vector representation}\label{S-ADAPT}
A major aspect of the numerical implementation is to establish a suitable consideration of polytopes and their dynamics. This is the purpose of this subsection. 

Since the affine part of $G_{D,\epsilon}$ is a multiple of the identity, the map $G_{D,\epsilon}$ itself preserves polytope facets orientation. A moment of reflection then concludes that the algorithm may deal exclusively with convex polytopes whose facets are aligned with discontinuity planes (because the corresponding set is invariant both under forward and backward dynamics). In particular, this is the case when the initial polytope is a {\bf cylinder set}, {\sl viz.}\ $P^0=A_{a_0\dots a_\ell}$ for some word $\{a_k\}_{k=0}^{\ell}$ with letters $a_k\in\{1,\dots,N_D\}$, where $A_{a_0\dots a_\ell}$ is defined by \cite{KH95} 
\[
A_{a_0\dots a_\ell}=\bigcap_{k=0}^{\ell}G_{D,\epsilon}^{-k}(A_{a_k}).
\]
In practice, candidate words $\{a_k\}_{k=0}^{\ell}$ for non-empty cylinders are obtained from empirical trajectories (whose OP is positive), by recording successive labels of visited atoms. 

Besides, following standards \cite{BV04}, polytopes in arbitrary dimension can be characterized as intersections of half-spaces, using inequality constraints on coordinates. Half-spaces in our context are delimited by discontinuity planes. Hence polytopes in the algorithm will be characterized using inequality constraints on the sums $\sum_{k=i}^{j}x_k$ of coordinates. 
Accordingly, given a vector 
\[
m=\left((\underline{m}_{i,j},\overline{m}_{i,j})_{i=1}^j\right)_{j=1}^D\in\R^{D(D+1)}\ \text{for which}\ \underline{m}_{i,j}<\overline{m}_{i,j}\ \forall i\leq j,
\]
we define the polytope $P_{m}$ by\footnote{Strict inequalities in this definition are chosen on purpose. Indeed, excluding polytope facets is a convenient way to exclude discontinuities from InAsUP construction.} 
\begin{equation}
P_m=\left\{x\in\R^D:\underline{m}_{i,j}<\sum_{k=i}^jx_k<\overline{m}_{i,j}, 1\leq i\leq j\leq D\right\},
\label{DEFPOLY}
\end{equation}
(provided that this set is not empty). Overdetermination in this expression is a convenient way to capture in a unique formal expression, all possible types of polytopes that can occur in the construction. In particular, while the number of atom facets may vary, every atom can be expressed as a polytope $P_m$ {\sl ie.}\ for every $A_a\subset S_D$, we have $A_a=P_{m_a}$ for some $m_a\in (\frac12+\Z)^{D(D+1)}$.

Not only polytopes can be represented by vectors (so that the polytope dynamics reduces to that of a $D(D+1)$-dimensional dynamical system), but some of their operations can be expressed in terms of vectors. The following operations are of special interest for our purpose. In all lines below, the indices $i,j$ run over all pairs such that $1\leq i\leq j\leq D$.
\begin{itemize}
\item[$\bullet$] Inclusion: $P_m\subset P_{m'}$ if $\underline{m'}_{i,j}\leq \underline{m}_{i,j}$ and $\overline{m}_{i,j}\leq \overline{m'}_{i,j}$. 
\item[$\bullet$] Intersection: 
$P_m\cap P_{m'}=P_{m\cap m'}$  where $\underline{(m\cap m')}_{i,j}=\max\{\underline{m}_{i,j},\underline{m'}_{i,j}\}$ and $\overline{(m\cap m')}_{i,j}=\min\{\overline{m}_{i,j},\overline{m'}_{i,j}\}$. 
\item[$\bullet$] Dilation: For every $a>0$, we have $a\text{Id}_{\T^D}(P_m)=P_{m'}$, where $\overline{\underline{m'}}_{i,j}=a\overline{\underline{m}}_{i,j}$. 
\item[$\bullet$] Translation: For every $x\in\R^D$, we have $P_m+x=P_{m'}$ where $\overline{\underline{m'}}_{i,j}=\overline{\underline{m}}_{i,j}+\sum_{k=i}^jx_k$. 
\item[$\bullet$] Symmetry: $-P_m=P_{\Sigma(m)}$ where $\underline{\Sigma(m)}_{i,j}=-\overline{m}_{i,j}$ and $\overline{\Sigma(m)}_{i,j}=-\underline{m}_{i,j}$. 
\end{itemize}
The union operation is however not so convenient. In particular, depending on $m$ and $m'$, there might not exist $m''\in \R^{D(D+1)}$ such that 
\[
P_m\cup P_{m'}=P_{m''}\ \text{mod}\ 0.
\] 
Therefore, when testing that $P^{t+1}\subset \bigcup_{k=0}^{t}P^k\ \text{mod}\ 0$, the algorithm can only test if for every $P_m\subset P^{t+1}$, there exists $P_{m'}$ in the collection $\bigcup_{k=0}^{t}P^k$ such that $P_m\subset P_{m'}$. 

While the intersection property ensures immediate detection of asymmetry failure, a consequence of this constrained inclusion test is that the algorithm may continue to run while InAsIUP construction has been completed. In particular, polytopes in the final union may overlap and the corresponding cardinality might be unnecessarily excessive.

Polytopes overlap can however be reduced by chopping off pieces. Indeed one can test if a polytope $P_m\subset P^{t+1}$ writes $P_m=\bigcup_\ell P_{m_\ell}\ \text{mod}\ 0$ with $P_{m_\ell}\subset P_{m'}\subset \bigcup_{k=0}^{t}P^k$ for some $\ell$, so that the construction can only retain the complementary $P_{m_\ell}$ that do not belong to any $P_{m'}$ in the existing union. Retaining smaller polytopes makes it more likely that the inclusion $P^{t+1}\subset \bigcup_{k=0}^{t}P^k\ \text{mod}\ 0$ holds for $t$ smaller, and thus that the final union cardinality is smaller. Such a diminution has been observed in practice, accompanied with substantial reduction of computer resources when $D>3$. 

For simplicity, we opted to test if $P_m\subset P^{t+1}$ decomposes as the union of two pieces, namely 
\[
P_m=P_{m_\text{in}}\cup P_{m_\text{out}}\ \text{mod}\ 0
\]
with $P_{m_\text{in}}=P_m\cap P_{m'}$ for some $P_{m'}\subset \bigcup_{k=0}^{t}P^k$ (so that it suffices to retain $P_{m_\text{out}}$), see Appendix \ref{A-CHOP} for mathematical foundations and criteria.\footnote{Actually, a more elaborated version of the algorithm also tests the three-piece decomposition
\[
P_m=P_{m_\text{in}}\cup P_{m_{\text{out},1}}\cup P_{m_{\text{out},2}}\ \text{mod}\ 0
\]
in the simplest case where $P_m$ intersects $P_{m'}$ transversally, so that the $P_{m_{\text{out},i}}$ result from chopping off along parallel faces.}

\subsubsection{Polytope constraint optimization}
\begin{figure}[h]
\begin{center}
\includegraphics*[height=35mm]{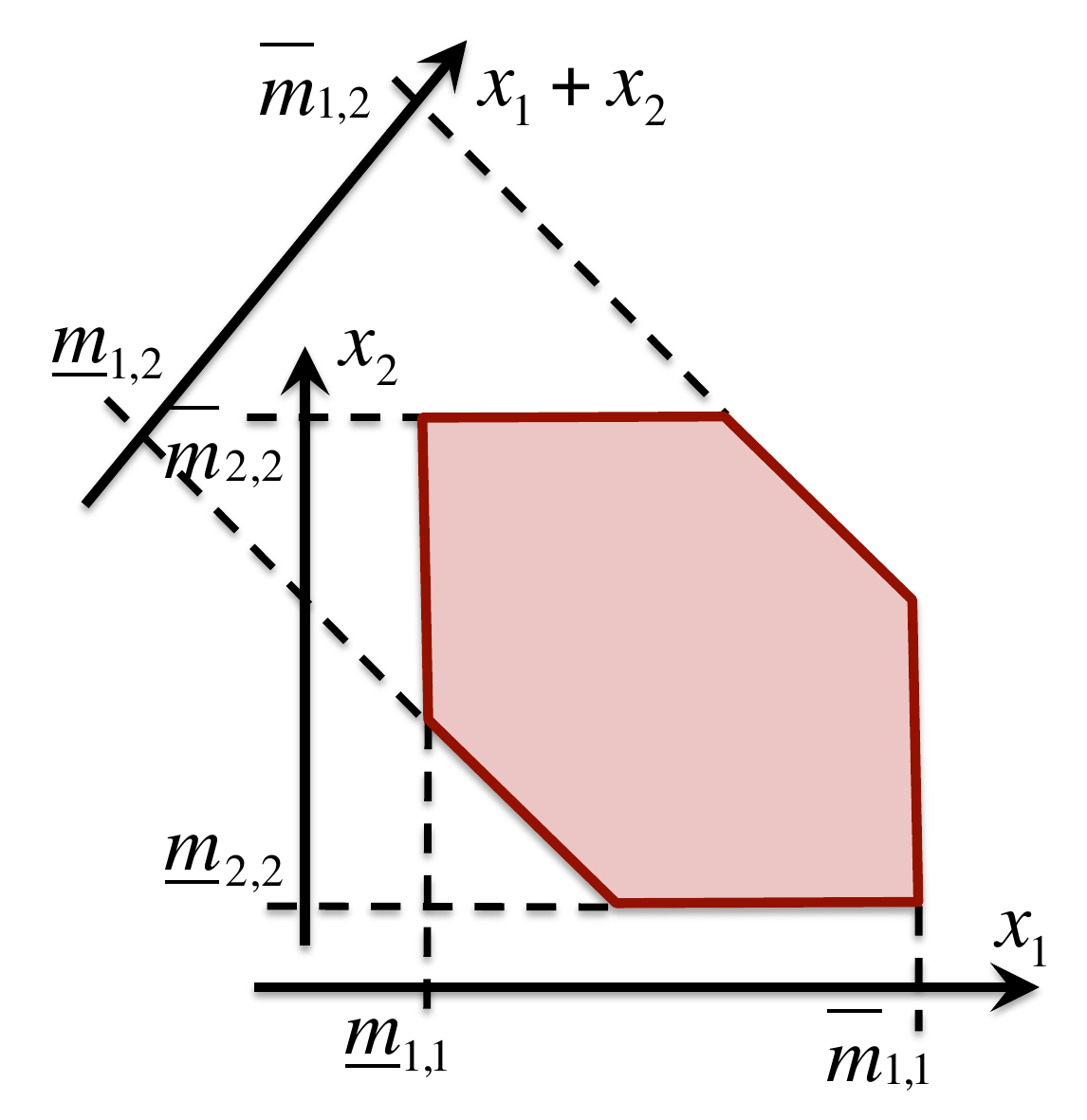}
\includegraphics*[height=35mm]{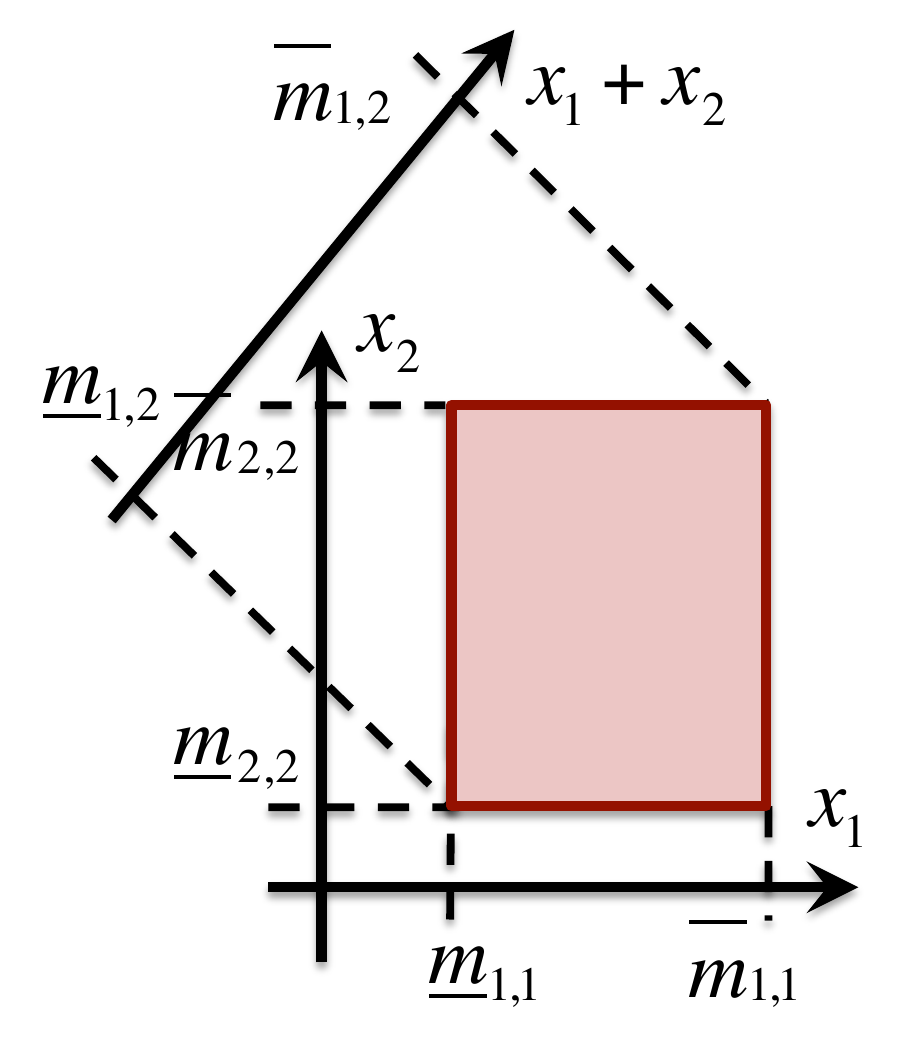}
\end{center}
\caption{Illustrations of polygons $P_m$ for $D=2$. {\sl Left.} Hexagonal example for which all constraints in expression \eqref{DEFPOLY} must be active. {\sl Right.} Rectangular example for which the constraints $\underline{m}_{1,2}\leq x_1+x_2\leq \overline{m}_{1,2}$ need not be active. Any vector $\left((\underline{m}_{i,j},\overline{m}_{i,j})_{i=1}^j\right)_{j=1}^2$ for which $\underline{m}_{1,2}\leq \underline{m}_{1,1}+\underline{m}_{2,2}$ and $\overline{m}_{1,1}+\overline{m}_{2,2}\leq \overline{m}_{1,2}$ defines the same $P_m$.}
\label{POLYTOPCONST}
\end{figure}

While expression \eqref{DEFPOLY} is convenient, it does not implies that $P_m$ is not empty, nor uniqueness of the constraining vector $m$. Indeed, $P_m$ can be unambiguously specified and yet, as many as $D(D-1)$ constraints may remain inactive (Fig.\ \ref{POLYTOPCONST}). Inactive constraints \cite{BV04} are problematic in an automated numerical implementation because they may yield spurious polytopes.

These issues call for constraint optimization. For the sake of notations, we present an optimization procedure in a slightly more general setting. The procedure relies on standard Linear Programming arguments. 

Let $D\leq E$ be arbitrary integers. Suppose that a collection $\alpha=\left((\alpha_{ij})_{j=1}^D\right)_{i=1}^E$ of distinct vectors with non-negative elements is given.\footnote{That is to say, $\alpha_{ij}\geq 0$ and $(\alpha_{ij})_{j=1}^D\neq (\alpha_{i'j})_{j=1}^D$ when $i\neq i'$.} For any constraint vector $m=(\underline{m}_i,\overline{m}_i)_{i=1}^{E}$, consider the polytope $P_m^\alpha$ defined by 
\[
P_m^\alpha=\left\{x\in \R^D\ :\ \underline{m}_i<\sum_{j=1}^D\alpha_{ij}x_j<\overline{m}_i,\ 1\leq i\leq E\right\}.
\]
For each $i\in\{1,\dots ,E\}$, consider the set $\Lambda_{i}$ of vectors $(\lambda_{k})_{k=1}^{E}$ with at least $E-D$ vanishing coordinates, 
which uniquely solve the system of equations
\begin{equation}
\sum_{k=1}^{E}\lambda_{k}\alpha_{kj}=\alpha_{ij},\ 1\leq j\leq D.
\label{LAGRANGE}
\end{equation}
More precisely, given any $s\in \{1,\dots,D\}$ and $S\subset\{1,\dots,E\}$ of cardinality $s$, if it exists, let $\lambda^{S}$ be the unique solution of the equations obtained from \eqref{LAGRANGE} by letting $\lambda_k=0$ for $k\in \{1,\dots,E\}\setminus S$. The set $\Lambda_i$ is made of all such solutions $\lambda^{S}$ when $S$ and $s$ vary.\footnote{In particular, $\Lambda_{i}$ is a finite set which can be obtained by listing all possible sets $S$ and computing, when they exist,  the unique solutions $\lambda^{S}$ of the corresponding systems. Moreover, $\Lambda_{i}$ cannot be empty because it contains the canonical vector $(\lambda_k)_{k=1}^{E}=(\delta_{k,i})_{k=1}^{E}$, where $\delta_{k,i}$ is the Kronecker symbol.}


Independently, given $\lambda\in\R$ and a constraint vector $m$, consider the vectors $(\underline{e}_k(\lambda,m))_{k=1}^{E}$ and $(\overline{e}_k(\lambda,m))_{k=1}^{E}$ respectively defined by 
\[
\underline{e}_{k}(\lambda,m)=\left\{\begin{array}{ccl}
\underline{m}_{k}&\text{if}&\lambda\geq 0\\
\overline{m}_{k}&\text{if}&\lambda< 0
\end{array}\right.
\quad\text{and}\quad
\overline{e}_{k}(\lambda,m)=\left\{\begin{array}{ccl}
\overline{m}_{k}&\text{if}&\lambda\geq 0\\
\underline{m}_{k}&\text{if}&\lambda< 0
\end{array}\right.
\]
Optimized constraint vectors, together with existence condition, are given in the following statement.
\begin{Lem}
Given any constraint vector $m=(\underline{m}_i,\overline{m}_i)_{i=1}^{E}$, consider $O(m)=(\underline{O(m)}_{i},\overline{O(m)}_{i})_{i=1}^{E}$ defined by
\begin{equation*}
\underline{O(m)}_i=\max_{(\lambda_k)\in\Lambda_i}\sum_{k=1}^{E}\lambda_k\underline{e}_k(\lambda_k,m)\quad\text{and}\quad
\overline{O(m)}_i=\min_{(\lambda_k)\in\Lambda_i}\sum_{k=1}^{E}\lambda_k\overline{e}_k(\lambda_k,m).
\end{equation*}
Then
\begin{itemize}
\item[(i)] $P_m^\alpha$ is not empty iff $\underline{O(m)}_i<\overline{O(m)}_i$ for all $1\leq i\leq E$.
\item[(ii)] If $P_m^\alpha$ is not empty, then $P_{O(m)}^\alpha=P_m^\alpha$ and all constraints in the definition of $P_{O(m)}^\alpha$ are active.
\item[(iii)] 
The plane $\sum_{j=1}^D\alpha_{ij}x_j=\underline{O(m)}_i$ (resp.\ $\sum_{j=1}^D\alpha_{ij}x_j=\overline{O(m)}_i$) defines a face of $P_m^\alpha$ iff 
\[
\max_{(\lambda_k)\in\Lambda_i\atop (\lambda_k)\neq (\delta_{k,i})}\sum_{k=1}^{E}\lambda_k\underline{e}_k(\lambda_k,O(m))<\underline{O(m)}_i\ 
\left(\text{resp.}\ \overline{O(m)}_i<\min_{(\lambda_k)\in\Lambda_i\atop (\lambda_k)\neq (\delta_{k,i})}\sum_{k=1}^{E}\lambda_k\overline{e}_k(\lambda_k,m).\right)
\]
\end{itemize}
\label{OPTIVEC}
\end{Lem}
The proof, given in Appendix \ref{A-OPTIM}, is inspired from the Karush-Kuhn-Tucker (KKT) approach to linear programming, an extension of the method of Lagrange multipliers to the case of inequalities constraints \cite{BV04}.

%
The algorithm extensively relies on Lemma \ref{OPTIVEC} and in particular, replaces $m$ by $O(m)$ every time the intersection of two polytopes is tested or computed.

\begin{table}
\begin{center}
\begin{tabular}{|c|c|c|c|c|c|}
\hline
D&2&3&4&5&6\\
\hline
$\#\Lambda_i$&2&5&16&65&326\\
\hline
\end{tabular}
\end{center}
\caption{Cardinality of the sets $\Lambda_i$ involved in the optimisation procedure for the collection $\alpha$ involved in \eqref{DEFPOLY}.}
\label{CARDICOMB}
\end{table}
For the collection $\alpha$ involved in the definition \eqref{DEFPOLY}, the cardinality of the sets $\Lambda_i$ does not depend on $i$ but it exponentially increases with $D$, see Table \ref{CARDICOMB}.  According to the GNU performance analysis tool {\tt gprof}, the optimization procedure $m\leftarrow O(m)$ is the most consuming resource task of the overall execution, in terms of CPU time. 

In order to speed up the process, the results in Table \ref{ALGOTAB} have actually been obtained using the semi-optimization procedure $O'$ which maximises/minimises over the subsets $\Lambda'_i\subset\Lambda_i$ of vectors having at most two non-vanishing coordinates. Naturally, the resulting vectors $O'(m)$ are no longer optimal (hence some of the $P_{O'(m)}$ may be spurious). Therefore, the cardinality of the constructing InAsIUP may be larger than when using $O$ only. However, this deficiency does not seem to affect InAsIUP successful completion in practice, especially because the chopping procedure of Appendix \ref{A-CHOP} still applies. More importantly, computation times appear to be reduced by a factor $\sim\tfrac{\#\Lambda_i}{\#\Lambda'_i}$ which, since $\#\Lambda'_i=D$, is a substantial gain for $D\geq 4$. 

\subsubsection{Polytope related vector dynamics}\label{S-POLDYN}
Let $B_a=B_D|_{A_a}$ be the expression in atom $A_a$ of the vector involved in the constant part of $G_{D,\epsilon}$. In each atom, $G_{D,\epsilon}$ regarded as acting on polytopes $P_m$ induces a mapping $\Gamma_{D,\epsilon,a}$ on vectors $m$, via the relation $G_{D,\epsilon}|_{A_a}(P_m)=P_{\Gamma_{D,\epsilon,a} (m)}$. Explicitly, we have 
\[
\overline{\underline{\Gamma_{D,\epsilon,a} (m)}}_{i,j}=2(1-\epsilon)\overline{\underline{m}}_{i,j}+\frac{2\epsilon}{D+1}\sum_{k=i}^jB_{a,k},\ \forall 1\leq i\leq j\leq D.
\]
Moreover, the atomic decomposition 
\[
P_m=\bigcup_{a}P_{m\cap m_a}\ \text{mod}\ 0,
\]
where $P_{m\cap m_a}=P_m\cap P_{m_a}$ and $P_{m_a}=A_a$, induces an extension of the vector map $\Gamma_{D,\epsilon}$ to arbitrary polytopes $P_m\subset S_D$, {\sl ie.}\ we have 
\[
\Gamma_{D,\epsilon} (m)=\bigcup_a\Gamma_{D,\epsilon,a} (m\cap m_a)
\] 
where the label index $a$ runs over all labels for which $P_{m\cap m_a}$ is not empty. Furthermore, in relation with the previous section, notice that the vector dynamics preserve the optimization procedure, {\sl ie.} we have 
\[
O\circ \Gamma_{D,\epsilon,a}=\Gamma_{D,\epsilon,a}\circ O.
\]

Notice finally that, since every $B_a\in\Q^D$, when $\epsilon\in\Q$ a rational number, every map $\Gamma_{D,\epsilon,a}$ has the properties 
\[
\Gamma_{D,\epsilon,a} (\Q^{D(D+1)})\subset \Q^{D(D+1)}\ \text{and}\ \Gamma_{D,\epsilon,a}^{-1} (\Q^{D(D+1)})\subset \Q^{D(D+1)},
\]
{\sl viz.}\ the polytope related vector dynamics effectively implies only rational numbers when $\epsilon\in Q$.

\section{Discussion}
The literature on the deterministic dynamics of collective systems mostly describes phenomenological changes related to pattern formation or transition to synchrony. These changes typically correspond to bifurcations of steady states or periodic solutions \cite{ABP-VRS05,DF18,CH93,PRK01}. These transitions -- which may also imply a reduction in the phase space to lower dimensional subspaces \cite{Buescu97} -- can be regarded as the breaking of ergodicity of an atomic or singular measure. Instead, analogues of phase transitions with spontaneous symmetry breaking should involve ACIM, in order to preserve all degrees of freedom of the corresponding chaotic attractor. Rigorously articulated examples of this are few, especially when appeal to a underlying {\sl ad hoc} random process possessing the desired property is excluded.

In this paper, we have first provided (complementary) numerical evidence for symmetry breaking of a chaotic attractor of full dimension, in a model of a collective system of interacting individuals (whose Markov partition remains elusive).  Furthermore, we have developed and benchmarked an algorithm for exact computer proof of the corresponding breaking of ACIM ergodicity. Even though the algorithm could only terminate for systems with limited number of individuals (due to limitations of computational resources), it indicates that phase transitions can be rigorously proven in a purely deterministic setting, without any reference to statistical mechanics. 

In order to improve their physical relevance, such features should be confirmed for systems with larger numbers of individuals and for more realistic models. For our systems, such confirmation might require algorithmic improvements. For instance, one may rely on parallel implementation -- even though the iterative construction of an invariant set is an intrinsically sequential process -- on automated decomposition into unions of non-overlapping polytopes, and on the use of optimized libraries for rational arithmetic, as well as dedicated computational resources.

Besides the existence of InAsUP, alternative proofs of the breaking of ACIM ergodicity could be obtained using spectral properties of the transfer operator associated with the coupled map system. Since this operator governs the dynamics of the densities associated with measures, it suffices to prove that it acquires multiple fixed points \cite{KL09} as the coupling intensity increases. Likewise, one could aim at delineating (coupling-dependent) Markov partitions, which are compatible with ergodicity at small coupling and simultaneously imply splitting into asymmetric ergodic components as interactions become strong. To our best knowledge, these approaches have not been considered in the literature.

In any case, while transitions in statistical mechanics only occur in the thermodynamic limit, $N\to +\infty$ -- especially because irreducible Markov chains, which govern the dynamics for finite $N$, must be uniquely ergodic -- this paper, together with \cite{F14,S18,SB16}, shows that without Markov chain considerations, deterministic models do not require to consider this limit, and can display similar non-ergodic behaviors and symmetry breaking in finite dimension. This feature is particularly interesting for the modelling of real particle systems, which must be finite-dimensional.

\subsection*{Acknowledgments}
I am grateful to P. B\'alint, J. Buzzi, V. Perchet, F. S\'elley and L-S. Young for scientific discussions, to Y. Legrandg\'erard, L. Ollivier and D. Simon for computational insights and to P.\ B\'alint, N.\ Cuneo and G.\ Francfort for a critical reading of the manuscript and thoughtful suggestions of improvements.

\appendix
\section{Symmetry breaking for $G_{2,\epsilon}$}\label{SymBreakG2}
The map $G_{2,\epsilon}$'s explicitly expression is given by $G_{2,\epsilon}=2(1-\epsilon)\text{\rm Id}_{\T^{2}}+\tfrac{2\epsilon}{3}B_2\ \text{mod}\ 1$, where
\[
\left\{\begin{array}{l}
(B_{2,\epsilon}(x))_1=2h(x_1)-h(x_2)+h(x_1+x_2)\\
(B_{2,\epsilon}(x))_2=2h(x_2)-h(x_1)+h(x_1+x_2)
\end{array}\right.
\]
$G_{2,\epsilon}$ commutes with each transformation in the natural representation of $\Z_2\times S_3$ on $\T^2$. This group can be generated by the sign flip $-\text{Id}_{\T^2}$ and the transformations $\sigma_{213},\ \sigma_{132}$ and $\sigma_{321}$, which are induced by the transpositions in $S_3$ (NB: subscripts here denote transposition images of the ordered list $\{1,2,3\}$ - abbreviated as 123). The sign flip writes
\[
- \text{Id}x = (-x_1,-x_2)\ (\text{mod}\ 1),
\]
and it corresponds to the central symmetry wrt $(\frac12,\frac12)$ in the square $(0,1)^2$. Similarly, we have
\[
\sigma_{321} x=(-x_2,-x_1)\ (\text{mod}\ 1),
\]
which corresponds to the orthogonal reflection wrt to the anti-diagonal $x_1+x_2=1$. Moreover, $\sigma_{213}$ (resp.\ $\sigma_{132}$) is defined by 
\[
\sigma_{213} x=(-x_1,x_1+x_2)\ (\text{mod}\ 1)\quad \left(\text{resp.}\ \sigma_{132} x=(x_1+x_2,-x_2)\ (\text{mod}\ 1)\right),
\]
is the reflection wrt the axis $x_1=0$ along the direction $x_1+2x_2=\text{cst}$ (resp.\ wrt the axis $x_2=0$ along the direction $2x_1+x_2=\text{cst}$). 

The representation on $\T^2$ of $\Z_2\times S_3$ itself can be listed as follows 
\begin{align*}
\{&\text{Id},-\text{Id},\ \sigma_{213},\ \sigma_{132},\ \sigma_{321},\ \sigma_{213}\circ \sigma_{321},\ \sigma_{321}\circ \sigma_{213},\ -\sigma_{213},\ -\sigma_{132},\ -\sigma_{321},\\
&-\sigma_{213}\circ \sigma_{321},\ -\sigma_{321}\circ \sigma_{213}\}.
\end{align*}
Clearly, all ergodic components for $\epsilon>\epsilon_2$ break the sign-flip symmetry (Fig.\ 1). More precisely, each component remains invariant under the action of one transformation above and is mapped onto another component under any other transformation. In particular, the chartreuse and dark green components are invariant under $\sigma_{213}$, Blue and light green components are invariant under $\sigma_{132}$ and red and orange components are invariant under $\sigma_{321}$.

\section{Symmetry breaking for $G_{3,\epsilon}$}\label{SymBreakG3}
The map $G_{3,\epsilon}$ writes $G_{3,\epsilon}=2(1-\epsilon)\text{\rm Id}_{\T^{3}}+\tfrac{\epsilon}{2}B_3\ \text{mod}\ 1$, where
\[
\left\{\begin{array}{l}
(B_{3,\epsilon}(x))_1=2h(x_1)-h(x_2)+h(x_1+x_2)-h(x_2+x_3)+h(x_1+x_2+x_3)\\
(B_{3,\epsilon}(x))_2=2h(x_2)-h(x_1)-h(x_3)+h(x_1+x_2)+h(x_2+x_3)\\
(B_{3,\epsilon}(x))_2=2h(x_3)-h(x_2)-h(x_1+x_2)+h(x_2+x_3)+h(x_1+x_2+x_3)
\end{array}\right.
\]
As before, any transformation in the representation of $S_4$ on $\T^3$ can be obtained as a composition of 2-element-permutation representations, whose expressions are given by (NB: the $(\text{mod}\ 1)$ are not specified for the sake of space)
\begin{align*}
&\sigma_{2134} x=(-x_1,x_1+x_2,x_3),&  &\sigma_{3214} x=(-x_2,-x_1,x_1+x_2+x_3)\\
&\sigma_{4231} x=(-x_2-x_3,x_2,-x_1-x_2),& &\sigma_{1324} x=(x_1+x_2,-x_2,x_2+x_3)\\
&\sigma_{1432} x=(x_1+x_2+x_3,-x_3,-x_2),& &\sigma_{1243} x=(x_1,x_2+x_3,-x_3)
\end{align*}

The ergodic components that emerge at $\epsilon=\epsilon_3$ all break the sign-flip symmetry $-\text{Id}$ in $\T^3$. As before, these 6 components are only partly asymmetric; they remain invariant under the action of 7 transformations in $\Z_2\times S_4$, and they are exchanged under other transformations. In particular, the blue component is invariant under 
\begin{align*}
&\sigma_{3214},\ \sigma_{1432},\ -\sigma_{1243}\circ \sigma_{2134},\ -\sigma_{1324}\circ \sigma_{4231},\ -\sigma_{1324}\circ \sigma_{1432}\circ \sigma_{2134}\\
&\text{and}\ -\sigma_{1243}\circ \sigma_{4231}\circ \sigma_{2134},
\end{align*}
and, obviously, under the composition $\sigma_{3214}\circ \sigma_{1432}=\sigma_{1432}\circ \sigma_{3214}$.

As for the 8 components emerging at $\epsilon \simeq 0.437$, their asymmetries are stronger than for the previous components, as they remain invariant under only 5 transformations. For instance, the fuschia component is invariant under 
\[
\sigma_{2134},\ \sigma_{4231},\ \sigma_{1432},
\]
and, obviously, under the compositions $\sigma_{4231}\circ \sigma_{2134}=\sigma_{2134}\circ \sigma_{4231}=\sigma_{2134}\circ \sigma_{1432}=\sigma_{1432}\circ \sigma_{4231}$ and $\sigma_{1432}\circ \sigma_{2134}=\sigma_{4231}\circ \sigma_{1432}$.

\section{Proof of Lemma \ref{OPTIVEC}}\label{A-OPTIM}
We first establish the expression of $\overline{O(m)}_i$. A similar analysis yields the expression of $\underline{O(m)}_i$. Let $i\in\{1,\dots,E\}$ and suppose that we aim to maximize the quantity $\sum_{j=1}^D\alpha_{ij}x_j$ given the $2E$ linear inequality constraints
\begin{equation}
\sum_{j=1}^D\alpha_{kj}x_j\leq \overline{m}_{k}\quad \text{and}\quad -\sum_{j=1}^D\alpha_{kj}x_j\leq -\underline{m}_{k}, 
\quad 1\leq k\leq E.
\label{LININCONS}
\end{equation}
This problem exactly fits the KKT approach to linear programming \cite{BV04}. The corresponding KKT conditions state in particular that every maximizer $(x_j^\ast)_{j=1}^D$ must cancel the gradient of the following Lagrangian
\[
\sum_{j=1}^D\alpha_{ij}x_j-\sum_{k=1}^{E}\gamma_k\left(\sum_{j=1}^D\alpha_{kj}x_j-\overline{m}_{k}\right)+\sum_{k=1}^{E}\gamma'_k\left(\sum_{j=1}^D\alpha_{kj}x_j-\underline{m}_{k}\right),
\]
for a unique pair of vectors $(\gamma_k)_{k=1}^{E},(\gamma'_k)_{k=1}^{E}$ with non-negative components. In other words, this pair of vectors must satisfy the equation
\begin{equation}
\sum_{k=1}^{E}(\gamma_k-\gamma'_k)\alpha_{kj}=\alpha_{ij},\ 1\leq j\leq D.
\label{LAG2}
\end{equation}
The complementary slackness conditions in the KKT setting then state that we must also have
\begin{equation}
\gamma_k\left(\sum_{j=1}^D\alpha_{kj}x_j^\ast-\overline{m}_{k}\right)=0\ \text{and}\ \gamma'_k\left(\sum_{j=1}^D\alpha_{kj}x_j^\ast-\underline{m}_{k}\right)=0,\ 1\leq k\leq E.
\label{KKTCOND}
\end{equation}
Now, the conditions $\underline{m}_{k}<\overline{m}_{k}$ imply that the constraints $\sum_{j=1}^D\alpha_{kj}x_j^\ast\leq \overline{m}_{k}$ and $-\sum_{j=1}^D\alpha_{kj}x_j^\ast\leq -\underline{m}_{k}$ cannot be simultaneously active, {\sl viz.}\ we must have $\gamma_k\gamma'_k=0$ for all $k$. A single multiplier $\lambda_k=\gamma_k-\gamma'_k$ results for each $k$ and equation \eqref{LAG2} implies that the vector $(\lambda_k)_{k=1}^{E}$ must solve equation \eqref{LAGRANGE}.

Equation \eqref{KKTCOND} also implies that for $\lambda_k\neq 0$ we have
\[
\sum_{j=1}^D\alpha_{kj}x_j^\ast=\left\{\begin{array}{ccl}
\overline{m}_k&\text{if}&\lambda_k=\gamma_k> 0\\
\underline{m}_k&\text{if}&\lambda_k=-\gamma'_k<0
\end{array}\right.=\overline{e}_{k}(\lambda_k,m).
\]
Together with equation \eqref{LAGRANGE}, this equality yields
\[
\sum_{j=1}^D\alpha_{ij}x_j^\ast=\sum_{k=1}^{E}\lambda_k\sum_{j=1}^D\alpha_{kj}x_j^\ast=\sum_{k=1}^{E}\lambda_k\overline{e}_{k}(\lambda_k,m),
\]
which is exactly the expression involved in the definition of $\overline{O(m)}_i$. 

When $E>D$, the system \eqref{LAGRANGE} is underdetermined. Hence it may have infinitely many solutions. The following considerations show that we may only retain solutions in $\Lambda_i$. 

Consider the partition of $\R^{E}$ into orthants in the interior of which the sign of every coordinate $\lambda_k$ is constant. Choose any orthant that contains a family of (infinitely many) solutions to \eqref{LAGRANGE} in its interior.\footnote{If no such orthant exists, then proceed with similar considerations in orthant boundaries, as indicated in the next paragraph.} Those coordinates $\lambda_k$ with identical sign ($E>2$) must have opposite variations when varying the solution in the family. However, the functional $(\lambda_k)_{k=1}^{E}\mapsto \sum_{k=1}^{E}\lambda_{k}\overline{e}_k(\lambda_k,m)$ is linear (and hence with constant gradient) inside the orthant. Therefore, it certainly reaches its maximum on the boundary of the orthant, {\sl ie.}\ when at least one coordinate of $(\lambda_k)_{k=1}^{E}$ vanishes. 

Repeating the reasoning inside orthant boundaries, which are identified with $\R^{E-1}$, and then for all subsequent subspaces $\R^{E'}$ for decreasing $E'$, from $E-2$ to $D+1$, {\sl viz.}\ as long as the resulting systems remain underdetermined, we conclude that we may only consider solutions of \eqref{LAGRANGE} with $E-D$ vanishing coordinates.

Yet, the system with $D$ unknowns/equations may also be degenerate. In this case, one can remove one superfluous equation from \eqref{LAGRANGE} and repeat the previous argument to conclude that the maximum must occur for a vector with $E-D+1$ vanishing coordinates. Repeating this reasoning as long as a degeneracy occurs, we may eventually get a system of 2 unknowns/equations. This system may be problematic if degenerate and the two coordinates have opposite signs. In fact, the only problematic scenario is if the functional $\sum_{k=1}^{E}\lambda_{k}\underline{e}_k(\lambda_k,m)$ increases when the positive coordinates increases (otherwise the maximum is reached when one coordinate vanishes). In this case, the maximum would be $+\infty$, which is certainly impossible. 
This analysis concludes that $\overline{O(m)}_i$ maximizes the quantity $\sum_{j=1}^D\alpha_{ij}x_j$, under the constraints \eqref{LININCONS}. 

As a consequence, if $P_m^\alpha\neq \emptyset$, then for every $x\in \overline{P_m^\alpha}$ (the closure of $P_m^\alpha$), we must have 
\[
\underline{O(m)}_i\leq \sum_{j=1}^D\alpha_{ij}x_j\leq \overline{O(m)}_i,\ \forall 1\leq i\leq E,
\]
{\sl viz.}\ $\overline{P_m^\alpha}\subset \overline{P_{O(m)}^\alpha}$ (and then $P_{O(m)}^\alpha\neq\emptyset$). On the other hand, that the canonical vector $(\delta_{k,i})_{k=1}^{E}$ belongs to $\Lambda_i$ implies
\[
\underline{m}_i\leq \underline{O(m)}_i\quad\text{and}\quad\ \overline{O(m)}_i\leq \overline{m}_i,\ \forall 1\leq i\leq E,
\]
hence $\overline{P_{O(m)}^\alpha}\subset \overline{P_m^\alpha}$. It results that $P_{O(m)}^\alpha=P_m^\alpha$ when $P_m^\alpha\neq \emptyset$.

Now, the complementary slackness conditions \eqref{KKTCOND} imply that the maximum $\overline{O(m)}_i$ and minimum $\underline{O(m)}_i$ are given by combinations of values at active constraints. Besides, we must have $\overline{O(m)}_i=\overline{m}_i$ and $\underline{O(m)}_i=\underline{m}_i$ when the constraints at $i$ are active, by the definitions of $\overline{O(m)}_i$ and $\underline{O(m)}_i$. It follows that the values of $\overline{O(m)}_i$ and $\underline{O(m)}_i$ do not change when we replace $m$ by $O(m)$ in the constraints \eqref{LININCONS} ({\sl viz.}\ $O$ is a projection operator) and the complementary slackness conditions imply that all constraints must be active in the updated optimisation problem, {\sl ie.}\ all constraints in the definition of $P_{O(m)}^\alpha$ must be active. 

Moreover, that $\overline{O(m)}_i$ and $\underline{O(m)}_i$ are coordinate extremal values in $P_{O(m)}^\alpha=P_m^\alpha$ immediately imply that 
\[
\underline{O(m)}_i< \overline{O(m)}_i,\ \forall 1\leq i\leq E,
\]
is an existence condition for $P_m^\alpha$. 

\noindent
{\sl Proof of item (iii).} If a maximizer $(x_j^\ast)_{j=1}^D$ satisfies 
\[
\sum_{j=1}^D\alpha_{ij}x_j^\ast=\overline{O(m)}_i<\min_{(\lambda_k)\in\Lambda_i\atop (\lambda_k)\neq (\delta_{k,i})}\sum_{k=1}^{E}\lambda_k\overline{e}_k(\lambda_k,m),
\]
then, by continuity, the inequality
\[
\sum_{j=1}^D\alpha_{ij}x_j<\min_{(\lambda_k)\in\Lambda_i\atop (\lambda_k)\neq (\delta_{k,i})}\sum_{k=1}^{E}\lambda_k\overline{e}_k(\lambda_k,m),
\]
holds for all points in the intersection of the plane $\sum_{j=1}^D\alpha_{ij}x_j=\overline{O(m)}_i$ with a sufficiently small neighborhood of  $(x_j^\ast)_{j=1}^D$; hence this plane defines a facet of $P_m^\alpha$. 

\noindent
On the opposite, if  
\[
\overline{O(m)}_i=\min_{(\lambda_k)\in\Lambda_i\atop (\lambda_k)\neq (\delta_{k,i})}\sum_{k=1}^{E}\lambda_k\overline{e}_k(\lambda_k,m),
\]
then let $\delta=(\delta_i)_{i=1}^D$ be a perturbation of the maximiser $x^\ast$ in the plane $\sum_{j=1}^D\alpha_{ij}x_j=\overline{O(m)}_i$, {\sl ie.}\ such that $\sum_{j=1}^D\alpha_{ij}\delta_j=0$. Then for those $(\lambda_k)$ for which the previous equality holds, we must also have
\[
\sum_{k=1}^{E}\lambda_k\sum_{j=1}^D\alpha_{kj}\delta_j=0\quad\Rightarrow\quad \sum_{j=1}^D\alpha_{kj}\delta_j=0,\ \forall k\ :\ \lambda_k\neq 0.
\]
Since the vectors $(\alpha_{kj})_{j=1^D}$ are all distinct, this means that $\delta$ must be limited to a set of co-dimension $1+\#\{k\ :\ \lambda_k\neq 0\}>1$. Therefore, it can certainly not span a facet of $P_m^\alpha$. The proof of Lemma \ref{A-OPTIM} is complete. \hfill $\Box$

\section{Mathematical statements related to polytope chopping tests}\label{A-CHOP}
The setting in this section assumes that two polytopes $P_m^\alpha$ and $P_{m'}^\alpha$ (where the constraint vectors $m$ and $m'$ are assumed to be optimized) intersect but are not contained in one another. We aim to determine (simple) conditions so that $P_m^\alpha\setminus P_{m\cap m'}^\alpha$ consists of either one or two polytopes defined by inequality constraints of the same type. We shall rely on the following set of indices
\[
S=\left\{1\leq i\leq E:\underline{m}_i<\underline{m'}_i\ \text{and/or}\ \overline{m'}_i<\overline{m}_i\right\}.
\]
\begin{Claim}
In the current setting, assume that\footnote{We have $i^-,i^+\in S$ and $i^-$ (resp.\ $i^+$) exists only when $\underline{m'}_{i^-}$ (resp.\ $\overline{m'}_{i^+}$) defines a facet of $P_{m'}^\alpha$.}
\begin{itemize}
\item a unique $i^-$ exists such that $\underline{m}_{i^-}<\underline{m'}_{i^-}$ and $\underline{O(m\cap m')}_{i^-}=\underline{m'}_{i^-}$ 
\item and/or, a unique $i^+$ exists such that $\overline{m'}_{i^+}<\overline{m}_{i^+}$ and $\overline{O(m\cap m')}_{i^+}=\overline{m'}_{i^+}$.
\end{itemize}
Then, the intersection $P_{m\cap m'}^\alpha:=P_m^\alpha\cap P_{m'}^\alpha$ can be characterized as follows
\[
\left(\underline{(m\cap m')}_i,\overline{(m\cap m')}_i\right)=\left\{\begin{array}{ccl}
(\underline{m}_i,\overline{m}_i)&\text{if}&i\not\in S\ \text{or}\ i\neq i^-,i^+\\
(\underline{m'}_{i^-},\overline{m}_{i^-})&\text{if}&i=i^-\neq i^+\\
(\underline{m}_{i^+},\overline{m'}_{i^+})&\text{if}&i=i^+\neq i^-\\
(\underline{m'}_i,\overline{m'}_i)&\text{if}&i=i^-=i^+
\end{array}\right.
\]
\end{Claim}
The numerical algorithm relies on the following consequence to define complementary piece(s) when an index of type $i^-$ or $i^+$ appears in testing intersections.
\begin{Cor}
If $i^-$ exists, let the constraint vector $m^-$ be defined by
\[
(\underline{m^-}_i,\overline{m^-}_i)=\left\{\begin{array}{ccl}
(\underline{m}_i,\overline{m}_i)&\text{if}&i\neq i^-\\
(\underline{m}_{i^-},\underline{m'}_{i^-})&\text{if}&i=i^-\end{array}\right.
\]
and similarly, let $m^+$ be defined by
\[
(\underline{m^+}_i,\overline{m^+}_i)=\left\{\begin{array}{ccl}
(\underline{m}_i,\overline{m}_i)&\text{if}&i\neq i^-\\
(\overline{m'}_{i^+},\overline{m}_{i^+})&\text{if}&i=i^+\end{array}\right.
\]
when assuming $i^+$ exists. We have  
\[
P_m^\alpha=P_{m^-}^\alpha\cup P_{m\cap m'}^\alpha\cup P_{m^+}^\alpha\ \text{mod}\ 0
\]
where $P_{m^-}^\alpha$ (resp.\ $P_{m^+}^\alpha$) has to be replaced by $\emptyset$ if $i^-$ (resp.\ $i^+$) does not exist.
\end{Cor}
The proof of the Corollary is immediate. The definitions of $m^-$ and $m^+$ immediately follow from the characterizations of $m\cap m'$ and $i^-,i^+$. That the corresponding polytopes are non-empty is a consequence of the fact that $m$ and $m'$ are optimized (so that there exists $x^-\in P_{m^-}^\alpha$ such that $x_{i^-}^-=\underline{m}_{i^-}$ and $x^+\in P_{m^+}^\alpha$ such that $x_{i^+}^+=\underline{m}_{i^+}$).

The Claim remains valid if one considers the semi-optimization procedure $O'$ (obtained by replacing the $\Lambda_i$ by $\Lambda'_i$). However $P_{m^-}^\alpha\neq\emptyset$ and $P_{m^+}^\alpha\neq \emptyset$ can no longer be granted. So in using $O'$ instead of $O$ when testing chopping, the algorithm may include spurious polytopes in the constructing collection. 

\noindent
{\sl Proof of the Claim.} Consider all possible cases of ordering of the coordinates of $m$ and $m'$ under the assumption $m\cap m'\neq\emptyset$. Then, from the definition of $m\cap m'$ in Section \ref{S-ADAPT}, we must have 
\[
\left(\underline{(m\cap m')}_i,\overline{(m\cap m')}_i\right)=\left\{\begin{array}{ccl}
(\underline{m}_i,\overline{m}_i)&\text{if}&i\not\in S\\
(\underline{m'}_{i},\overline{m}_{i})&\text{if}&\underline{m}_i<\underline{m'}_i<\overline{m}_i\leq \overline{m'}_i\\
(\underline{m}_{i},\overline{m'}_{i})&\text{if}&\underline{m'}_i\leq \underline{m}_i<\overline{m'}_i<\overline{m}_i\\
(\underline{m'}_i,\overline{m'}_i)&\text{if}&\underline{m}_i< \underline{m'}_i<\overline{m'}_i<\overline{m}_i
\end{array}\right.
\]
Assume now that $i\in S$, otherwise there is nothing to prove. Consider first the case $\underline{m}_i<\underline{m'}_i<\overline{m}_i\leq \overline{m'}_i$. Then we must have $i\neq i^+$ and we can let $\overline{(m\cap m')}_i=\overline{m}_i$. So,
\begin{itemize}
\item either $i=i^-$ and we get $\underline{(m\cap m')}_{i^-}=\underline{m'}_{i^-}$ from the definition of $i^-$.
\item or $i\neq i^-$ and then we must have $\underline{m}_{i}<\underline{m'}_{i}<\underline{O(m\cap m')}_{i}$. Obviously, $\underline{m'}_{i}$ cannot define an active constraint of $P_{m\cap m'}^\alpha$ and we may set $\underline{(m\cap m')}_{i}=\underline{m}_{i}$ without affecting this set.
\end{itemize}
Now, the second case $\underline{m'}_i\leq \underline{m}_i<\overline{m'}_i<\overline{m}_i$ can be treated similarly. The third case is even simpler because either $i=i^-=i^+$ and then we get $(\underline{m'}_i,\overline{m'}_i)$, or $i\in\{i^-, i^+\}$ and $i^+\neq i^-$ and we get one of the previous cases. \hfill $\Box$

\section{Algorithm for InAsUP construction}\label{A-ALGO}
{\bf Declarations, initialisations}
\begin{algorithmic}
\STATE Choose $D,\epsilon$
\STATE Import vectors $(\lambda_k)_{k=1}^{\frac{D(D+1)}{2}}\in\bigcup_{i=1}^D\Lambda_i$ from output of automated Mathematica code.
\STATE  Import symbolic word $a_0\cdots a_\ell$ of initial cylinder, from symbolic sequence of empirical trajectory simulated using Mathematica code.
\end{algorithmic}

\noindent
{\bf Preliminaries: Computations of atom constraint vectors $m_a$ and constant part of $\Gamma_{D,\epsilon,a}$}
\begin{algorithmic}
\STATE Set $a \leftarrow 0$
\STATE  List all possible vectors $\left\{h\left(\sum_{k=i}^jx_k\right)\right\}_{1\leq i\leq j\leq D}$ for $x\in S_D$.
\FOR{each vector}
\STATE $m\leftarrow \left((\underline{m}_{i,j},\overline{m}_{i,j})_{i=1}^j\right)_{j=1}^D$ where $\underline{m}_{i,j}=h\left(\sum_{k=i}^jx_k\right)-\frac12$ and $\overline{m}_{i,j}=h\left(\sum_{k=i}^jx_k\right)+\frac12$
\STATE Apply optimization, {\sl ie.}\ $m\leftarrow O(m)$
\STATE Store constraint vector, {\sl ie.}\ $m_a\leftarrow m$
\STATE Store constant part vector, {\sl ie.}\ $B_a\leftarrow\frac{2\epsilon}{D+1}\left\{\sum_{k=i}^jB_{a,k}\right\}_{1\leq i\leq j\leq D}$
\STATE Set $a\leftarrow a+1$
\ENDFOR
\end{algorithmic}

\noindent
{\bf Implementation}
\begin{algorithmic}
\STATE Replace $O$ by the semi-optimized procedure $O'$, to be used in all future computations.\footnote{For the sake of brevity, optimizations $m\leftarrow O'(m)$ are not indicated in the rest of the algorithm. However, they are applied every time the intersection of two polytopes is tested or computed and when chopping into pieces is tested or computed.}
\STATE Compute constraint vector $m_{a_0\cdots a_\ell}$ of initial cylinder recursively, {\sl ie.}\ $m_{a_0\cdots a_\ell}=m_{a_0}\cap \Gamma_{D,\epsilon,a_0}^{-1}(m_{a_1\cdots a_\ell})$.
\STATE Set $\text{UP}\leftarrow m_{a_0\cdots a_\ell}$ and $\text{NP}\leftarrow m_{a_0\cdots a_\ell}$. 
\IF{$m_{a_0\cdots a_\ell}\cap \Sigma(m_{a_0\cdots a_\ell})\neq\emptyset$}
\STATE Exit implementation ({\bf Asymmetry fails})
\ENDIF
\WHILE{$\text{NP}\neq \emptyset$ $\&$ asymmetry holds $\&$ $\#\text{UP}<\text{MaxCard}$ {\bf (Main Loop)}}
\FOR{each $m\in \text{NP}$}
\STATE Compute image $m'=\Gamma_{D,\epsilon,a}(m)$.
\STATE Apply projection $\T^D\to S_D$. May chop $m'$ into several pieces (when $m'$ spans across boundaries of $S_D$). 
\FOR{each piece of $m'$ and each symbol $a$}
\IF{$m'\cap m_a\neq \emptyset$}
\STATE Set $m'\leftarrow m'\cap m_a$
\IF{$m'\cap \Sigma(\text{UP})\neq\emptyset$ $\parallel$ $m'\cap \Sigma(\text{EP})\neq\emptyset$}
\STATE Exit implementation ({\bf Asymmetry fails})
\ELSE
\FOR{each $m''\in (\text{UP}\cup \text{EP})\cap m_a$\footnote{In practice, this loop consists of two distinct loops; one for $m''\in \text{UP}\cap m_a$ followed by one for $m''\in \text{EP}\cap m_a$.}}
\IF{$m'\cap m''\neq\emptyset$ but $m'\not\subset m''$}
\IF{$m'=m^-\cup m''\cup m^+$ ($m^-\neq\emptyset$ $\parallel$ $m^+\neq\emptyset$)}
\STATE $m'\leftarrow (m^-\cup m^+)$ ({\bf Chopping operation}).
\ENDIF
\ENDIF
\ENDFOR
\STATE Set $\text{EP}\leftarrow \text{EP}\cup m'$.\footnote{Due to chopping, $m'$ may consists of several elements.}
\ENDIF
\ENDIF
\ENDFOR
\ENDFOR
\STATE Set $\text{NP}\leftarrow \text{EP}$, $\text{UP}\leftarrow \text{UP}\cup \text{NP}$ and $\text{EP}\leftarrow\emptyset$.
\ENDWHILE
\PRINT $\# \text{UP}$ and CPU time to complete implementation.
\end{algorithmic}

\noindent
{\bf Annex: Definitions of functions}
\begin{algorithmic}
\STATE Optimization function $m\leftarrow O(m)$ (and semi-optimization function $m\leftarrow O'(m)$).
\STATE Mapping $\Gamma_{D,\epsilon,a}$ in each atom.
\STATE Pre-image $\Gamma_{D,\epsilon,a}^{-1}$ in each atom.
\STATE Compute optimized intersection $O(m\cap m')$
\STATE Test $m\cap m'\neq 0$
\STATE Compute symmetric image $\Sigma(m)$.
\STATE Test chopping $m$ by $m'$ into one or two polytope(s).
\STATE Compute chopped pieces $m=m^-\cup (m'\cap m)\cup m^+$
\end{algorithmic}
\end{document}